\documentclass{raa} 
\usepackage{graphicx,times}
\usepackage{natbib}
\usepackage{amssymb,amsmath}
\usepackage{ulem}
\usepackage{color}
\usepackage{pdflscape}
\usepackage{graphicx,amsmath,amssymb,color,appendix,soul}

\newcommand{\kms}{\mbox{$\mathrm{km\,s^{-1}}$}}
\newcommand{\Ha}{\mbox{${\mathrm H\alpha}$}}
\newcommand{\Hbeta}{\mbox{${\mathrm H\beta}$}}
\newcommand{\Ion}[2]{#1{\,\scriptsize #2}}

\newcommand{\Lin}[3]{\Ion{$[$#1}{#2}$]$\,$\lambda$\,#3}
\newcommand{\Li}[3]{\Ion{$[$#1}{#2}$]$\,#3}
\newcommand{\Lines}[3]{\Ion{$[$#1}{#2}$]$\,$\lambda\lambda$\,#3}
\newcommand{\LinesNa}[3]{\Ion{#1}{#2}\,$\lambda\lambda$\,#3}

\begin{document}

    \title{On the radial velocity calibrations in the LAMOST Medium-Resolution Spectroscopic Survey of Nebulae}

 \volnopage{ {\bf 2012} Vol.\ {\bf X} No. {\bf XX}, 000--000}
   \setcounter{page}{1}

\author{Juan-Juan Ren\inst{1,3}, Hong Wu\inst{2,3}, Chao-Jian Wu\inst{2,3}, Wei Zhang\inst{2,3}, Jian-Jun Chen\inst{2,3}, Chih-Hao Hsia\inst{4}, Fan Yang\inst{2,3}, Chao Liu\inst{1,3,5}, Jian-Rong Shi\inst{2,3,5}, Yu-Zhong Wu\inst{2,3}, Hui Zhu\inst{3}, Bin Li\inst{6,7}, Zhong-Rui Bai\inst{2,3}, Hao Tian\inst{1,3}, and Yong-Hui Hou\inst{8,9}}

\institute{CAS Key Laboratory of Space Astronomy and Technology, National Astronomical Observatories, Chinese Academy of Sciences, Beijing 100101, China; {\it jjren@nao.cas.cn}\\
\and
CAS Key Laboratory of Optical Astronomy, National Astronomical Observatories, Chinese Academy of Sciences, Beijing 100101, China\\
\and
National Astronomical Observatories, Chinese Academy of Sciences, 20A Datun Road, Chaoyang District, Beijing 100101, China\\
\and
State Key Laboratory of Lunar and Planetary Sciences, Macau University of Science and Technology, Taipa, Macau, China\\
\and
University of Chinese Academy of Sciences, 100049, China\\
\and
Purple Mountain Observatory, Chinese Academy of Sciences, Nanjing 210008, China
\and
University of Science and Technology of China, Hefei 230026, China
\and
Nanjing Institute of Astronomical Optics, \& Technology, National Astronomical Observatories, Chinese Academy of Sciences, Nanjing 210042, China \\
\and
School of Astronomy and Space Science, University of Chinese Academy of Sciences \\
\vs \no
}

\abstract{
Accurate radial velocity determinations of optical emission lines (i.e. \Lines{N}{II}{6548,\,6584}, \Ha, and \Lines{S}{II}{6717,\,6731}) are very important for investigating the kinematics and dynamics properties of nebulae. The second stage survey program of Large sky Area Multi-Object fiber Spectroscopic Telescope (LAMOST) has started a sub-survey of nebulae (MRS-N) which will spectroscopically observe the optical emission lines of a large sample of nebulae near the Galactic plane. Until now, 15 MRS-N plates have been observed from 2017 September to 2019 June. Based on fitting the sky emission lines in the red band spectra of MRS-N, we investigate the precision of wavelength calibration and find there are systematic deviations of radial velocities (RVs) from $\sim$\,0.2 to 4\,\kms\ for different plates. Especially for the plates obtained in 2018 March, the systematic deviations of RVs can be as large as $\sim$\,4\kms, which then go down to  $\sim$\,0.2\,--\,0.5\,\kms at the end of 2018 and January 2019. A RVs calibration function is proposed for these MRS-N plates, which can simultaneously and successfully calibration the systematic deviations and improve the precision of RVs. 
}
\authorrunning{J.-J. Ren et al.}
\titlerunning{LAMOST MRS-N RV}
\maketitle

\section{Introduction}

Nebulae are interstellar clouds consisting of dust, hydrogen, helium and other ionized gases. Among the various types of nebulae, emission nebulae are clouds of interstellar ionized gases and dust that emit light in various wavelengths. There are different kinds of emission nebulae, including \Ion{H}{II} regions, planetary nebulae (PNe), supernovae remnants \citep[SNRs;][]{Woltjer1972ARA&A..10..129W} and so on. Different types of emission nebulae have various different formation mechanisms, thus they can be used to study many aspects in the astronomical sciences from the star formation to the evolution of Milky Way. \Ion{H}{II} regions can be used as probes of the composition of the interstellar medium in galaxy, and their emission-line spectrum can be used to determine the gas-phase abundance of several elements \citep{Shaver1983MNRAS.204...53S, Esteban2019arXiv190510129E}. Due to the bright emission lines of PNe, they can be used as tracers of galaxy kinematics \citep{Merrett2006MNRAS.369..120M},  and as potential tracers of the chemical evolution of Milky Way and other galaxies \citep{Cavichia2017MNRAS.468..272C, Kwitter2012ApJ...753...12K}. SNRs can provide insights into the mechanisms of supernova explosions, and probe the immediate surroundings of supernovae. SNRs are also fundamentally related to the star-forming process in a galaxy, and can also give us a picture of the on-going massive star formation rate \citep{Kopsacheili2020MNRAS.491..889K}.

Optical spectroscopic observations of emission nebulae can provide important information of nebulae, especially the \Ha, \Hbeta, \Lines{N}{II}{6548,\,6584}, and \Lines{S}{II}{6717,\,6731}, \Lines{O}{II}{3726,\,3729}, \Lines{O}{III}{4959,\,5007}. These optical emission lines can be used to investigate the kinematics and dynamics properties of nebulae \citep{Damiani2016A&A...591A..74D}. Furthermore, the optical line intensity ratio \Lin{S}{II}{6717}/\Lin{S}{II}{6731} is electron density sensitive \citep{Osterbrock2006agna.book.....O}, \Ha/\Lines{S}{II}{6717,6731} ratio can be used to distinguish shocked nebulae from photoionized nebulae \citep{Alvarez2005A&A...440..569A}, and the \Ha/\Li{N}{II} ratio is a widely used tool for investigating nitrogen-to-hydrogen abundance variations among SNRs \citep{Fesen1985ApJ...292...29F}. The emission line ratios combining with the kinematics of ionized gas, can be used to characterize and identify the nature of the nebulae. Thus it is very important to obtain complete sampling of the optical line emissions in various nebulae. A lot of optical spectrophotometric observations have been carried out for emission nebulae. But most of them are only dedicated for some special nebulae \citep{Fesen1985ApJ...292...29F, Blair1991ApJ...366..484B, Gerardy2007MNRAS.376..929G, Chen2017MNRAS.472.3924C}, thus still lack of spectroscopic survey to sample a large sample of nebulae near the Galactic plane, especially for those with large spacial size. 

The Large sky Area Multi-Object fiber Spectroscopic Telescope (LAMOST) is a quasi-meridian reflecting Schmidt telescope, which has a $\sim$4 meters effective aperture and a field of view of 5$^\circ$ diameter \citep{Wang1996ApOpt..35.5155W, Su2004ChJAA...4....1S, Cui2012RAA....12.1197C, Zhao2012RAA....12..723Z}. As a dedicated spectroscopic survey telescope, LAMOST can acquires optical spectra with 4000 fibers during one single exposure. Since September 2012, LAMOST has been carrying out the first five-years Regular Surveys, i.e. the first stage survey program of LAMOST (hereafter LAMOST\,I). The LAMOST\,I Regular Surveys mainly consist of two components \citep{Zhao2012RAA....12..723Z}: the LAMOST Extra-Galactic Survey of Galaxies (LEGAS) that aims to study the large scale structure of the Universe; and the LAMOST Experiment for Galactic Understanding and Exploration \cite[LEGUE;][]{Deng2012RAA....12..735D} with the goal of obtaining millions of stellar spectra to study the structure and evolution of the Milky Way. The LAMOST\,I spectra cover the entire optical wavelength range ($\simeq$\,3700\,--\,9000\,\AA), with a resolving power R\,$\sim$\,1800 \citep{Luo2012RAA....12.1243L, Luo2015RAA....15.1095L}. 

Since September 2017, the second stage survey program of LAMOST (hereafter LAMOST\,II) has been initiated. Except continuing the low-resolution spectroscopic observation, LAMOST\,II began the medium-resolution spectroscopic survey \citep[MRS;][]{Liu2020arXiv200507210L}. In every lunar month, half of the nights (i.e. the dark/gray nights, from 23rd to 6th on lunar calendar) are used to continue the previous low-resolution spectroscopic surveys, while the remaining half of the nights (i.e. the bright/gray nights, from 7th to 22nd on lunar calendar) are assigned for the MRS. LAMOST\,II MRS spectra have a medium-resolution of R\,$\sim$\,7500, and cover the wavelength ranges of 4950\,--\,5350\,\AA\ in blue channel and 6300\,--\,6800\,\AA\ in red channel \citep{Wu2020RAA....20...33W}. The main scientific aims of LAMOST\,II MRS include the time-domain (TD) and non-time-domain (NT) sciences. Hence the MRS survey can be broken down into a few different sub-surveys: the Kepler region survey (MRS-K), the TESS follow-up survey (MRS-T), the Star forming region survey (MRS-S), the Binary survey (MRS-B), the Galactic Nebulae survey (MRS-N), the Galactic archaeology survey (MRS-G), and Open cluster survey (MRS-O). Among them, MRS-K, MRS-T, MRS-B, and part fields of MRS-S are assigned as the TD survey, while the MRS-G, MRS-N, MRS-O, and part of the MRS-S are operated as NT survey \citep{Liu2020arXiv200507210L}. 

The large field of view (5$^\circ$), relatively higher resolution (7500), and dedicated wavelength coverage (6300\,--\,6800\,\AA, thus covering the emission lines \Ha, \Lines{N}{II}{6548,\,6584}, and \Lines{S}{II}{6717,\,6731}) make LAMOST\,II MRS an ideal spectroscopic survey to sample the optical emission lines of Galactic nebulae \citep{Wu2020arXiv200705240W} in unprecedented details. To better investigate the kinematics properties of nebulae, one of the most important things is to obtain better wavelength calibration thus improving the radial velocity determination of nebulae emission lines. 

In this paper, we investigate the wavelength calibration and finally propose a method to improve the precision of the radial velocity determinations in the LAMOST\,II MRS nebulae survey. In Section 2, we describe the observation and data reduction of LAMOST\,II MRS nebulae survey. In Section 3 we present the method, and Section 4 presents the discussion. Finally Section 5 gives a summary. 

\begin{figure}[htbp]
    \centering
	\includegraphics[angle=0,width=150mm]{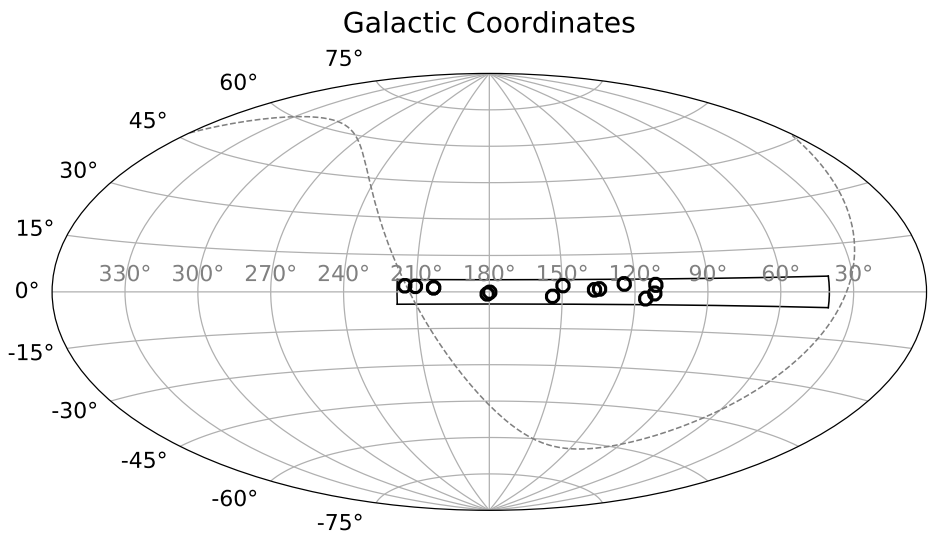}
    \caption{The Galactic coordinates of the 15 MRS-N plates (black open circles). The black square shows the survey region of MRS-N, i.e. 40$^\circ$\,$<$\,l\,$<$\,218$^\circ$ and $-$5$^\circ$\,$<$\,b\,$<$\,5$^\circ$.}
    \label{fig:obs_plates}
\end{figure}

\section{LAMOST\,II MRS-N observation and data reduction}
\subsection{Observation}
LAMOST\,II MRS nebulae (MRS-N) survey aims at obtaining the medium-resolution spectra of a large sample of Galactic nebulae including the \Ion{H}{II} regions, PNe, SNRs, and Herbig-Haro objects \citep{Wu2020arXiv200705240W}. The MRS-N survey contains two components: (1) obtain the optical spectra of emission lines of a large sample of Galactic nebulae in Galactic plane (40$^\circ$\,$<$\,l\,$<$\,218$^\circ$, $-$5$^\circ$\,$<$\,b\,$<$\,5$^\circ$, $\sim$\,1700 deg$^2$ totally), with every field only one time observation; (2) obtain the optical spectra of four specially selected nebulae: Westerhout 5 (HII regions), Rosette Nebula\,+\,NGC2264 (HII region), Cygnus Loop (SNRs) and Simeis 147 (SNRs), each one with more than 10 observations during the five-years regular survey. As the moonlight has serious influence of the nebulae observation, which makes the sky subtraction very difficult for nebulae spectra, thus the MRS-N observations only were carried out during the moonless gray/bright nights. LAMOST\,II MRS-N survey was initiated in September 2017. The first year of the survey was a pilot survey, and the first regular survey started since September 2018. 

Until June 2019, 15 MRS-N plates have been observed. Table\,\ref{tab:obs_plates} lists the detailed information of these 15 MRS-N plates. Figure\,\ref{fig:obs_plates} shows their Galactic distribution. From Table\,\ref{tab:obs_plates} and Figure\,\ref{fig:obs_plates}, we can see that most of the observed MRS-N plates are concentrated near the Galactic anti-center, with the Galactic longitude in the range 110$^\circ$\,--\,215$^\circ$, and the Galactic latitude between $-$2.7$^\circ$ and 3.2$^\circ$. Most of the MRS-N plates have exposure time 900s $\times$ 3, which is relatively short than the exposure time of the normal star observations (1200s $\times$ 3) of MRS. There are two reasons: (1) the very limited observation time for nebulae observation, which should be carried out during the moonless gray/bright nights; (2) another reason should be that the 900s $\times$ 3 exposure is already enough to obtain good quality of emission line spectra of nebulae. 

\begin{table*}
	\centering
	\caption{The 15 plates observed by LAMOST\,II MRS-N in the pilot the first year regular survey (from September 2017 to June 2019). Here we list the observation date (ObsDate), the plate ID, the name of the center star, the coordinates of the center star, the seeing during the observation, and the exposure time (Exp).}
	\label{tab:obs_plates}
	\setlength{\tabcolsep}{2pt}
	\begin{tabular}{ccccccccc} 
		\hline
		 ObsDate	& Plate	& Center Star &	RA	&	Dec	& l &  b  & Seeing  &	Exp \\
		            &       &    & ($^\circ$) &  ($^\circ$) & ($^\circ$) &  ($^\circ$) &   (arcsec)   &   (s) \\
		\hline
20180305 & HD474690101        & HD47469 & 99.7966 &    9.64596  & 202.956762 &    1.691792 & 5.6 & 900s $\times$ 3  \\
20180305 & HD474690801        & HD47469 & 99.7966 &    9.64596  & 202.956762 &    1.691792 & 4.8 & 900s $\times$ 3  \\
20181018 & NT054437N290059N01 & HIP27088 & 86.1573 &   29.0166   & 179.818302 & $-$0.142857 & 2.4 & 1200s $\times$ 3 \\
20181118 & NT070341S003325N01 & HIP34039 & 105.9240 & $-$0.556953 & 214.826342 &    2.472767 & 3.4 & 900s $\times$ 3 \\
20181128 & NT230608N631246N01 & HIP114070 & 346.5370 &   63.2128   & 111.391030 &    2.724941 & 2.9 & 900s $\times$ 3 \\
20181128 & NT235302N592517N01 & HIP117772 & 358.2600 &   59.4216   & 115.557042 & $-$2.611379 & 2.9 & 900s $\times$ 3 \\
20181129 & NT010552N655815N01 & HIP115228 &  16.4707 &   65.9711   & 124.403798 &    3.140640 & 3.7 & 900s $\times$ 3  \\
20181129 & NT232020N601629N01 & HIP5147 & 350.0870 &   60.2748   & 111.866024 & $-$0.625150 & 3.6 & 900s $\times$ 3 \\
20181216 & NT065450N031415N01 & HIP33227 & 103.7110 &  3.23735  & 210.437214 &    2.232564 & 5.0 & 900s $\times$ 3 \\
20181228 & NT023239N614403N01 & HD15557 & 38.1664 &   61.7342   & 134.613668 &    1.180085 & 5.3 & 900s $\times$ 3 \\
20190124 & NT024709N603414N01 & HD17086 & 41.7903 &   60.5708   & 136.683682 &    0.826460 & 2.7 & 900s $\times$ 3, 1200s $\times$ 2 \\
20190125 & NT041422N543127N01 & HIP19774 & 63.5933 &   54.5242   & 149.660595 &    2.546033 & 2.8 & 900s $\times$ 3 \\
20190125 & NT041453N482433N01 & HIP19812 & 63.7244 &   48.4093   & 153.939273 & $-$1.824453 & 2.9 & 900s $\times$ 3 \\
20190126 & NT054421N274353N01 & HD38084 & 86.0882 &   27.7315   & 180.881876 & $-$0.866799 & 3.5 & 900s $\times$ 3 \\
20190126 & NT054421N274353N02 & HD38084 & 86.0882 &   27.7315   & 180.881876 & $-$0.866799 & 3.7 & 900s $\times$ 2 \\
		\hline
		\end{tabular}
\end{table*}

\subsection{Data Reduction}

\begin{figure}
    \centering
	\includegraphics[angle=0,width=150mm]{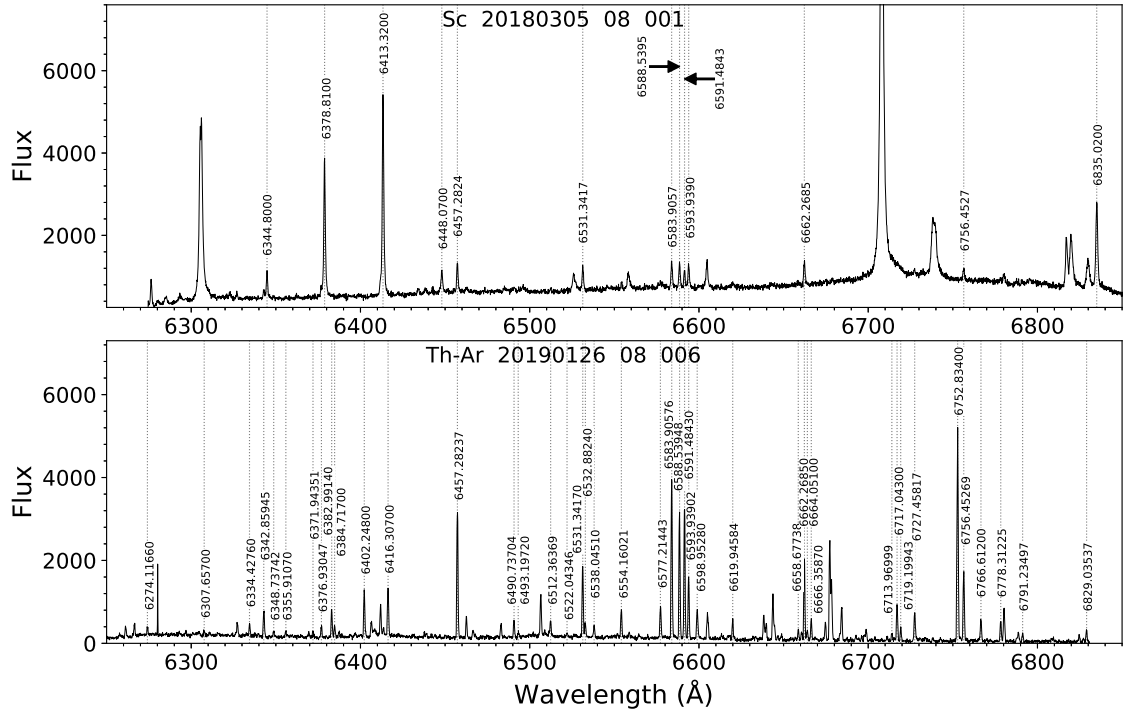}
    \caption{An example Sc (upper panel) and Th-Ar (bottom panel) arc lamp spectrum in red band. The vertical gray dotted lines mark the arc lines used for MRS wavelength calibration.}
    \label{fig:arc_lines}
\end{figure}

As the MRS-N observations and scientific goals are very different from other normal MRS surveys, there will be a new dedicated data reduction pipeline for MRS-N survey (Wu et al., 2020c, in preparation). The LAMOST MRS raw spectra are processed with the LAMOST two-dimensional (2D) pipeline \citep{Luo2015RAA....15.1095L}, including the dark and bias subtraction, cosmic ray removal, one-dimensional (1D) spectral extraction, and wavelength calibration. The MRS wavelength calibration procedures are similar to those for the LAMOST low-resolution spectra reduction pipeline (as described in \citet{Luo2015RAA....15.1095L}) by using arc lines. We need to note that the Sc lamps were used before May 2018, and then Th-Ar lamps were used instead. After the extraction of arc lamp spectra, the centroids of the arc lines are measured, which will be used to fit a Legendre polynomial (fifth-order in the blue channel, and sixth-order in the red) as a function to describe the relationship between wavelengths and pixels. Then the wavelength will be calibrated to vacuum, and also corrected to the heliocentric frame \citep{Luo2015RAA....15.1095L}. Usually about 42 Th-Ar arc lines or 13 Sc arc lines (only for the data obtained before May 2018) are used for the wavelength calibration of MRS spectra (see Figure\,\ref{fig:arc_lines} for an example of Sc and Th-Ar arc lamp spectrum). For different fibers, the number of arc lines used will be a little different due to their possible slightly different wavelength coverage. The spectra the MRS-N pipeline used are those extracted 1D sub-spectra, which are wavelength calibrated, but without sky subtraction and flux calibration. 

Then the MRS-N pipeline will merge the sub-spectra (usually three sub-exposures for each plates), re-calibrate the wavelength (the work we will investigate and propose in this paper), do the sky subtraction (a multi-component fitting method developed for MRS-N survey), fitting the emission lines in the red band (i.e. \Ha, \Lines{N}{II}{6548,\,6584}, and \Lines{S}{II}{6717,\,6731}), and finally publish the value-added catalogue for MRS-N survey, which includes the emission line radial velocities, line widths, and line intensity ratios (as presented in \citet{Ren2018RAA....18..111R}). Mostly, about 50\%--80\% of the \Li{N}{II} and \Li{S}{II} has the spectral S/N\,$>$\,10, and 80\%--95\% of \Ha\, can have S/N\,$>$\,10, depending on the sky regions observed (Wu et al. 2020c, in preparation).  

\begin{figure}
    \centering
	\includegraphics[angle=0,width=150mm]{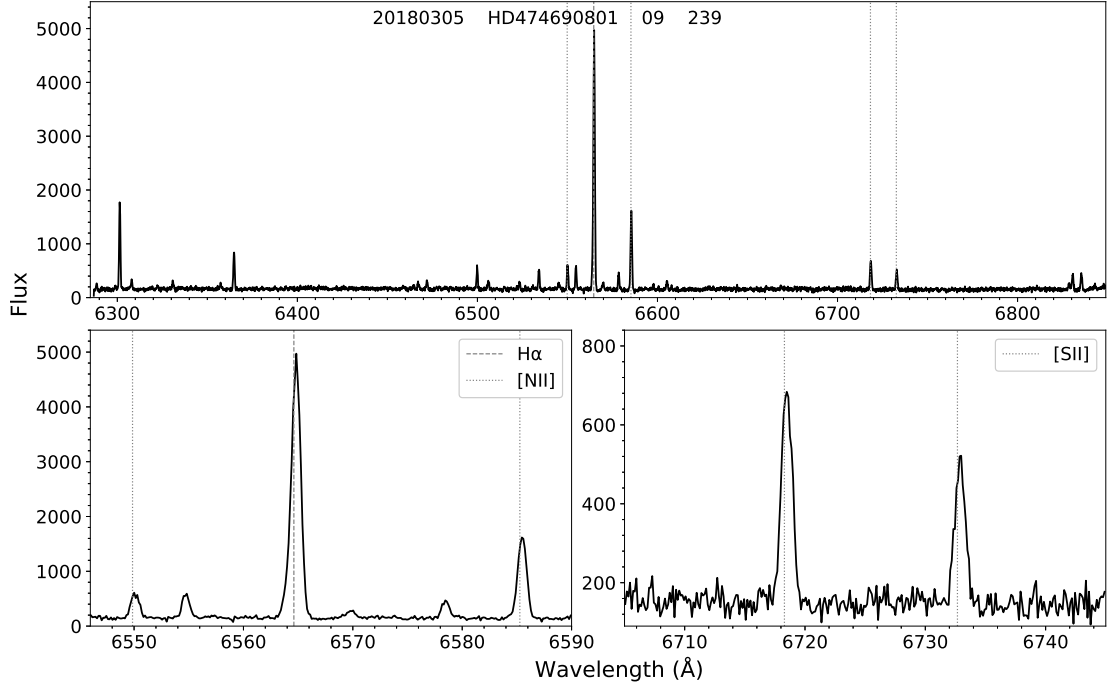}
    \caption{An example MRS spectrum of nebulae in red band (spectrum ID "20180305-HD474690801-09-239"). The bottom panels show the zoomed regions near the main emission features \Ha, \Lines{N}{II}{6548,\,6584}, and \Lines{S}{II}{6717,\,6731}.}
    \label{fig:spec_example}
\end{figure}

Here we need to note that, the LAMOST\,II MRS spectra have the wavelength coverage: 4950\,--\,5350\,\AA\ in blue and 6300\,--\,6800\,\AA\ in red channel. The blue spectra of MRS can not cover the important nebulae emission features in blue band (i.e. \Hbeta, \Lines{O}{II}{3726,3729}). Only \Lines{O}{III}{4959,5007} is available in the MRS blue spectra, while it is usually not as strong and visible as the emission features in the red band (\Ha, \Lines{N}{II}{6548,\,6584}, and \Lines{S}{II}{6717,\,6731}). Thus finally the MRS-N data reduction will only focus on the red band, and the final value-added MRS-N catalogue only provide the information determined from the red spectra too. Figure\,\ref{fig:spec_example} shows an example spectrum of MRS-N in red band.

\section{Method}

Accurate Radial Velocity (RV) determinations of emission lines are very important for studying the kinematics and dynamics properties of nebulae. While for the MRS observations, the arcs (Th-Ar) used for wavelength calibration are obtained before or after the normal observations (i.e. not simultaneously). Thus the wavelength calibration from arcs may incorporate some uncertainties for RV determinations, due to direction change of the telescope or environment variation (such as temperature, pressure) in the spectrograph room. To obtain accurate RV measurements, it's necessary to do a calibration. Unlike the RV calibrations of stars which can be obtained by using the RV standard stars \citep{Liu2019RAA....19...75L, Wang2019ApJS..244...27W}, the main targets in MRS-N are nebulae features not stars. Thus new method should be developed to investigate the wavelength calibration precision and propose new RV calibration ways to improve the RV precision in MRS-N survey data.

For the LAMOST low-resolution spectra reduction, some strong sky emission lines were used during the wavelength calibration procedure. But for the MRS spectra, the sky emission lines were not used in the 2D data reduction pipeline. As we mentioned in Section 1, the MRS observations were carried out during the bright/gray nights (from 7th to 22nd on lunar calendar) every month. That means, the moonlight background is very bright during the MRS observations for most times, and the corresponding sky emission lines have very low S/N, thus are not available for the wavelength calibration. While for our MRS-N survey, the observations are scheduled only during the moonless bright/gray nights, thus the sky emission lines still have good quality, and can be used to help us to improve the MRS-N wavelength calibration.

\begin{table*}
	\centering
	\caption{The obvious 7 single sky emission lines in the red band spectra of MRS-N. We list the name, wavelength in air and vacuum.}
	\label{tab:sky_lines_single}
	\setlength{\tabcolsep}{2pt}
	\begin{tabular}{cccc} 
		\hline
		 N &  Name	& Wavelength (air) & Wavelength (vacuum) \\
		   &        &            (\AA) & (\AA)  \\
		\hline
  1 & $\lambda$6287 & 6287.434 & 6289.1730 \\
  2 & $\lambda$6300 & 6300.304 & 6302.0464 \\
  3 & $\lambda$6363 & 6363.780 & 6365.5395 \\
  4 & $\lambda$6498 & 6498.729 & 6500.5248 \\
  5 & $\lambda$6533 & 6533.044 & 6534.8490 \\
  6 & $\lambda$6544 & 6544.022 & 6545.8299 \\
  7 & $\lambda$6553 & 6553.617 & 6555.4275 \\
		\hline
		\end{tabular}
\end{table*}

\begin{table*}
	\centering
	\caption{Similar to Table\,\ref{tab:sky_lines_single}, but for 6 blended sky emission lines.}
	\label{tab:sky_lines_blend}
	\setlength{\tabcolsep}{2pt}
	\begin{tabular}{cccc} 
		\hline
		 N &  Name & Wavelength (air) & Wavelength (vacuum) \\
		   &       &            (\AA) & (\AA)  \\
		\hline
  1 & $\lambda$6306 & [6306.869, 6306.981] & [6308.6132, 6308.7252] \\
  2 & $\lambda$6329 & [6329.747, 6329.933] & [6331.4974, 6331.6834] \\
  3 & $\lambda$6356 & [6356.167, 6356.441] & [6357.9244, 6358.1985] \\
  4 & $\lambda$6604 & [6603.990, 6604.279] & [6605.8140, 6606.1031] \\
  5 & $\lambda$6829 & [6827.459, 6828.469, 6829.491, 6829.564, 6829.922] & [6829.3432, 6830.3534, 6831.3757, 6831.4487, 6831.8068] \\
  6 & $\lambda$6834 & [6834.008, 6834.433] & [6835.8939, 6836.3190] \\
		\hline
		\end{tabular}
\end{table*}

\begin{figure}
    \centering
	\includegraphics[angle=0,width=150mm]{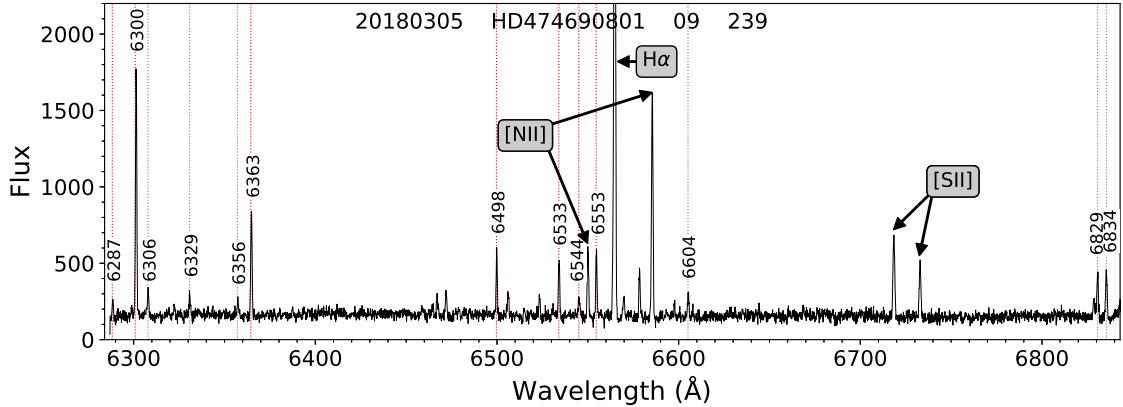}
    \caption{An example MRS-N spectrum showing the sky emission lines in red band (spectrum ID ``20180305-HD474690801-09-239"). The red dotted lines show the single lines listed in Table\,\ref{tab:sky_lines_single}, while the gray dotted lines show the blended lines  presented in Table\,\ref{tab:sky_lines_blend}.}
    \label{fig:sky_lines}
\end{figure}

\begin{figure}
    \centering
	\includegraphics[angle=0,width=155mm]{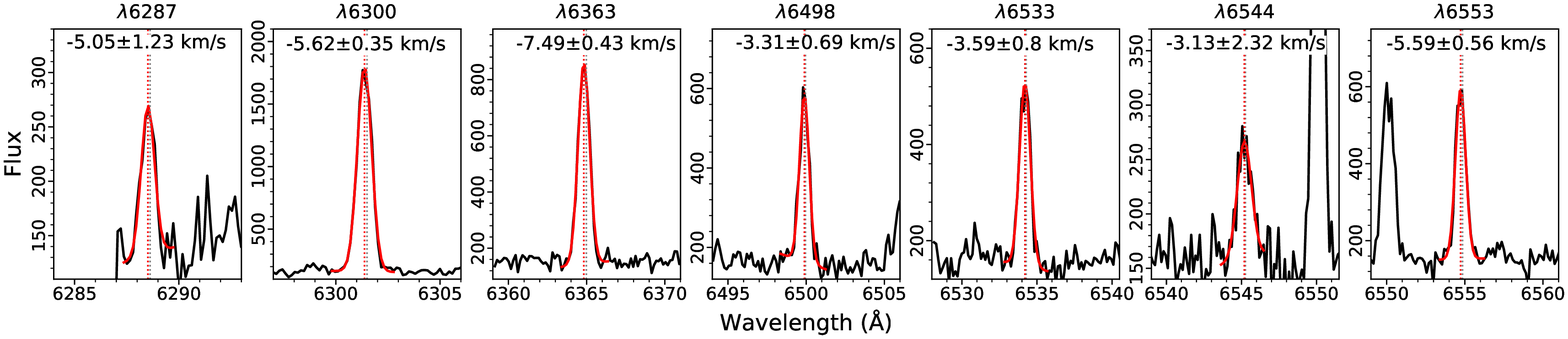}
    \caption{An example of single-Gaussian fitting of the sky emission lines (spectrum ID ``20180305-HD474690801-09-239") for 7 single lines. The red curves show the Gaussian fitted results, and the red dotted lines show the fitted centers. The gray dotted lines are the true line centers as listed in Table\,\ref{tab:sky_lines_single}.}
    \label{fig:fit_sky_lines_single}
\end{figure}

In the MRS-N 1D wavelength calibrated spectra without sky subtraction, there are  several sky emission lines as shown in Figure\,\ref{fig:spec_example}. There are 13 obvious sky emission lines in the red band \citep{Osterbrock1996PASP..108..277O}, of which 7 are single lines as listed in Table\,\ref{tab:sky_lines_single} and 6 are blended lines as presented in Table\,\ref{tab:sky_lines_blend}. Figure\,\ref{fig:sky_lines} presents an example MRS-N spectra showing these sky emission lines. Initially we hope to use all these sky emission lines (7 single + 6 blend lines) to calibrate RVs. By carefully investigating the blending situation and line profiles of the blended lines, we find it's hard to differentiate the multi-peaks of the blended lines, thus it's impossible to obtain accurate line centers even by multi-gaussian fitting under our relatively low resolution ($\sim$\,7500) spectra. Much higher resolution ($\sim$\,30\,000) spectra are needed to well differentiate the multi-peaks of the blended lines \citep{Osterbrock1996PASP..108..277O}. If we apply single gaussian fitting to these blended lines, the fitted line centers may have relatively large uncertainties due to their blending features. Hence finally we exclude these 6 blended lines from our following analysis, and only use the 7 single lines. These sky emission lines cover the a relatively large wavelength range (6290\,--\,6560\,\AA) of the MRS-N red spectra (see Figure\,\ref{fig:sky_lines}), and their intrinsic wavelengths are already well-known and fixed (see Table\,\ref{tab:sky_lines_single}), thus they are ideal tools to investigate the precision of wavelength calibration of MRS in real time. Furthermore, the sky emission lines are obtained at the same time as the nebulae emission lines, so most importantly they can provide a simultaneous RV calibration and finally help us improve the precision of RV determinations of MRS-N.

\begin{table*}
	\centering
	\caption{The fitted RVs of 7 single sky emission lines for the spectra from the 15 MRS-N plates. `$-$' marks those the RVs are unavailable. Here we just list part of the table, the complete table can be found online.}
	\scriptsize
	\label{tab:sky_RVs_each_spec}
	\setlength{\tabcolsep}{1.5pt}
	\begin{tabular}{crrrrrrr} 
		\hline
		 SpecID & RV$_{6287}$ & RV$_{6300}$ & RV$_{6363}$ & RV$_{6498}$ & RV$_{6533}$ & RV$_{6544}$ & RV$_{6553}$   \\
		        & (\kms) & (\kms) & (\kms) & (\kms) & (\kms) & (\kms) & (\kms) \\
		\hline
20190126-NT054421N274353N01-02-120 & $-$4.68$\pm$1.33 & $-$1.77$\pm$0.46 & $-$0.19$\pm$0.46 & $-$0.66$\pm$0.58 & $-$1.26$\pm$1.05 & $-$3.87$\pm$1.64 &    0.07$\pm$0.98 \\
20190126-NT054421N274353N01-02-121 & $-$4.96$\pm$1.65 & $-$1.64$\pm$0.48 & $-$0.28$\pm$0.56 & $-$2.36$\pm$1.02 & $-$1.85$\pm$1.16 & $-$6.26$\pm$2.04 & $-$2.95$\pm$0.84 \\
20190126-NT054421N274353N01-02-122 & $-$4.37$\pm$1.85 & $-$0.10$\pm$0.30 &    1.56$\pm$0.40 &    0.27$\pm$0.82 &    1.16$\pm$0.65 &    4.75$\pm$0.71 &    0.70$\pm$1.44 \\
20190126-NT054421N274353N01-02-123 & $-$5.68$\pm$1.97 & $-$1.46$\pm$0.54 &    0.71$\pm$0.75 & $-$0.69$\pm$0.73 & $-$1.25$\pm$0.78 & $-$1.84$\pm$1.40 & $-$2.58$\pm$0.93 \\
20190126-NT054421N274353N01-02-124 &    0.93$\pm$1.79 & $-$0.06$\pm$0.46 &    0.29$\pm$0.61 &    0.14$\pm$0.50 & $-$1.18$\pm$0.89 & $-$6.52$\pm$2.39 & $-$2.22$\pm$0.84 \\
20190126-NT054421N274353N01-02-125 & $-$6.53$\pm$1.66 & $-$0.80$\pm$0.35 & $-$1.51$\pm$0.63 &    1.72$\pm$0.72 &    0.78$\pm$1.16 & $-$2.02$\pm$2.51 & $-$2.11$\pm$0.71 \\
20190126-NT054421N274353N01-02-126 & $-$5.26$\pm$1.11 &    0.68$\pm$0.30 &    0.51$\pm$0.30 & $-$0.95$\pm$0.81 & $-$0.44$\pm$1.19 & $-$0.05$\pm$2.63 & $-$0.69$\pm$0.74 \\
20190126-NT054421N274353N01-02-127 &    0.15$\pm$2.80 &    0.81$\pm$0.41 &    2.42$\pm$0.63 & $-$1.52$\pm$0.67 & $-$1.65$\pm$0.62 & $-$0.42$\pm$2.01 &    0.18$\pm$0.52 \\
20190126-NT054421N274353N01-02-128 & $-$7.03$\pm$1.40 &    1.18$\pm$0.27 &    0.93$\pm$0.53 &    0.60$\pm$0.84 &    0.30$\pm$0.96 &    1.76$\pm$1.24 & $-$1.02$\pm$0.99 \\
20190126-NT054421N274353N01-02-129 &    2.54$\pm$2.58 &    1.28$\pm$0.42 &    1.37$\pm$0.56 & $-$1.28$\pm$0.73 & $-$1.86$\pm$1.30 & $-$1.85$\pm$2.38 & $-$0.43$\pm$0.91 \\
20190126-NT054421N274353N01-02-130 & $-$1.95$\pm$1.17 &    0.27$\pm$0.67 &    0.81$\pm$0.52 & $-$0.08$\pm$0.98 & $-$0.75$\pm$0.79 & $-$6.32$\pm$2.51 & $-$1.29$\pm$1.15 \\
		\hline
		\end{tabular}
\end{table*}

Gaussian fitting method is widely used for measuring the stellar RVs based on SDSS \citep{Rebassa2016MNRAS.458.3808R} or LAMOST spectra \citep{Ren2018MNRAS.477.4641R}. Usually a a second-order polynomial plus a single-Gaussian line profile was used to fit a single line (such as H$\alpha$ emission line), while a second-order polynomial plus a double/triple-Gaussian line profile of
fixed separation was used to fit the double/triple lines (like \LinesNa{Na}{I}{8183.27,\,8194.81} or \Ion{Ca}{II} absorption triplet at 8498.02, 8542.09, and 8662.14\,\AA), as described in \citet{Rebassa2007MNRAS.382.1377R, Rebassa2017MNRAS.472.4193R} and \citet{Ren2013AJ....146...82R, Ren2014A&A...570A.107R}. Here we adopt the similar method, i.e. a second-order polynomial plus a single-Gaussian line profile is used to fit the 7 single sky emission lines in Table\,\ref{tab:sky_lines_single}. Figure\,\ref{fig:fit_sky_lines_single} shows an example of the Gaussian fitting results of one nebula spectrum. 

Table\,\ref{tab:sky_RVs_each_spec} lists the fitting results for every spectrum of the 15 MRS-N plates. For those strong sky emission lines like $\lambda$6300 and $\lambda$6363 (as shown in Figure\,\ref{fig:sky_lines} and \ref{fig:fit_sky_lines_single}), the fitting error is only $\lesssim$\,0.4\,--\,0.5\,\kms; for those not so strong lines like $\lambda$6498, $\lambda$6533 and $\lambda$6553, the fitting error is $\lesssim$\,0.7\,--\,1.0\,\kms ; even for those relatively weak emission lines (i.e. $\lambda$6287 and $\lambda$6544), the fitting error is $\lesssim$\,1.6\,--\,2\,\kms mostly, all of which shows the good quality of sky emission lines, thus makes them the best tools to investigate and improve the precision of RV determinations of nebulae emission lines. From Figure\,\ref{fig:fit_sky_lines_single}, we can see that there is obvious systematic deviation around 4\,\kms\ for the RV of this spectrum. 

\begin{figure}
    \centering
	\includegraphics[angle=0,width=155mm]{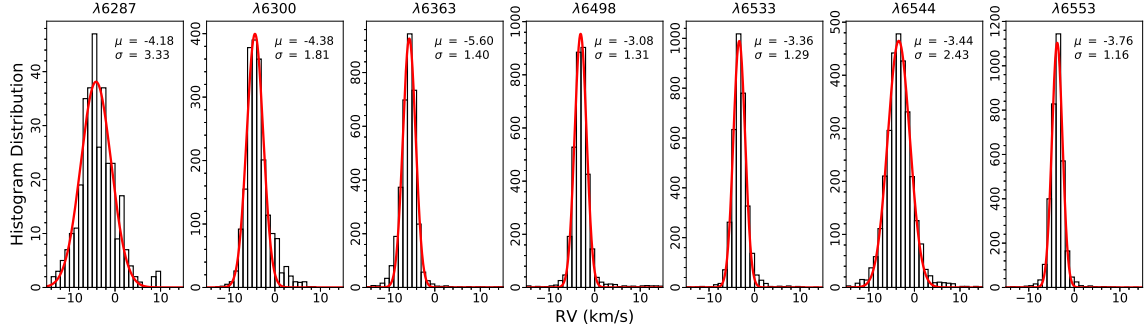}
    \caption{The histogram distribution of the RVs of 7 single sky emission lines for MRS-N plate "20180305-HD474690801".The red curves are the Gaussian fitting results of the histogram distributions, the $\mu$ and $\sigma$ marked in each panel are the center and standard deviation of the fitting.}
    \label{fig:hist_sky_RVs}
\end{figure}

To investigate the situation of the whole plate, we obtain the mean and standard deviation of the fitted sky RVs for every MRS-N plate (see Table\,\ref{tab:rv_hist_results}), by gaussian-fitting of the histogram distribution of sky RVs. Figure\,\ref{fig:hist_sky_RVs} shows the histogram distribution of the fitted RVs for the 7 single sky emission lines for plate ``20180305-HD474690801". For those relatively strong sky emission lines (i.e. $\lambda$6300, $\lambda$6363, $\lambda$6498, $\lambda$6533, $\lambda$6553), the standard deviation is lower ($\sim$\,1.2\,--\,2.0\,\kms); while for those relatively weak lines (such as $\lambda$6287, $\lambda$6544), the standard deviation is a little large ($\sim$\,2.4\,--\,3.4\,\kms), which can be easily explained as the weak lines have low spectral quality thus can transfer relatively large fitted uncertainties. From Figure\,\ref{fig:hist_sky_RVs} and Table\,\ref{tab:rv_hist_results}, we can clearly see the systematic deviation of wavelength calibration for almost the whole plate, thus it's very necessary to calibration these systematic deviations to obtain accurate RVs of nebulae emission lines. 

\begin{table*}
	\centering
	\caption{The gaussian-fitted mean and standard deviation of the histogram distribution of sky RVs for 15 MRS-N plates.}
	\scriptsize
	\label{tab:rv_hist_results}
	\setlength{\tabcolsep}{3pt}
	\begin{tabular}{ccrrrrrrr} 
		\hline
		 ObsDate & plateID & $\mu_{6287}$ & $\mu_{6300}$ & $\mu_{6363}$ & $\mu_{6498}$ & $\mu_{6533}$ & $\mu_{6544}$ & $\mu_{6553}$ \\
		         &         &       (\kms) &       (\kms) &       (\kms) &       (\kms) &    (\kms)    &  (\kms)      &    (\kms)    \\
		\hline
20180305 & HD474690101        &  $-$4.16$\pm$3.23 &  $-$4.56$\pm$1.71  &  $-$5.76$\pm$1.40  &  $-$3.24$\pm$1.34  &  $-$3.50$\pm$1.35  &  $-$3.50$\pm$2.71  &  $-$3.99$\pm$1.23  \\
20180305 & HD474690801        &  $-$4.18$\pm$3.33 &  $-$4.38$\pm$1.81  &  $-$5.60$\pm$1.40  &  $-$3.08$\pm$1.31  &  $-$3.36$\pm$1.29  &  $-$3.44$\pm$2.43  &  $-$3.76$\pm$1.16  \\
20181018 & NT054437N290059N01 &  $-$              &  $-$7.67$\pm$2.54  &  $-$4.52$\pm$1.46  &  $-$1.76$\pm$1.32  &  $-$1.96$\pm$1.39  &  $-$1.73$\pm$2.39  &  $-$2.19$\pm$1.21  \\
20181118 & NT070341S003325N01 &  $-$              &  $-$               &     0.82$\pm$1.33  &     0.71$\pm$1.21  &     0.21$\pm$1.27  &     0.34$\pm$1.70  &  $-$0.03$\pm$1.16  \\
20181128 & NT230608N631246N01 &  $-$              &  $-$               &  $-$0.41$\pm$2.26  &  $-$0.35$\pm$2.26  &  $-$0.25$\pm$2.19  &     0.02$\pm$4.01  &  $-$0.52$\pm$2.06  \\
20181128 & NT235302N592517N01 &  $-$              &  $-$               &  $-$0.06$\pm$1.97  &  $-$0.18$\pm$2.07  &  $-$0.05$\pm$2.02  &     0.14$\pm$3.27  &  $-$0.30$\pm$1.92  \\
20181129 & NT010552N655815N01 &  $-$              &  $-$               &     0.80$\pm$1.81  &     0.01$\pm$2.27  &     0.11$\pm$2.27  &     0.55$\pm$3.88  &     0.20$\pm$2.16  \\
20181129 & NT232020N601629N01 &  $-$              &  $-$               &     0.45$\pm$2.03  &     0.02$\pm$2.20  &     0.16$\pm$2.17  &     0.33$\pm$3.46  &     0.09$\pm$2.01  \\
20181216 & NT065450N031415N01 &  $-$0.62$\pm$2.69 &     0.51$\pm$1.24  &     0.63$\pm$1.44  &     0.29$\pm$1.30  &     0.08$\pm$1.39  &     0.08$\pm$2.65  &  $-$0.28$\pm$1.28  \\
20181228 & NT023239N614403N01 &  $-$0.61$\pm$2.93 &     0.42$\pm$1.91  &     0.67$\pm$1.97  &     0.63$\pm$1.76  &     0.17$\pm$1.75  &     0.30$\pm$2.79  &  $-$0.02$\pm$1.65  \\
20190124 & NT024709N603414N01 &  $-$1.36$\pm$3.37 &  $-$0.40$\pm$1.58  &  $-$0.32$\pm$1.51  &  $-$0.58$\pm$1.61  &  $-$0.70$\pm$1.73  &  $-$0.31$\pm$3.35  &  $-$0.95$\pm$1.57  \\
20190125 & NT041422N543127N01 &  $-$0.51$\pm$3.53 &     0.42$\pm$1.48  &     0.44$\pm$1.44  &  $-$0.06$\pm$1.84  &  $-$0.42$\pm$1.90  &  $-$0.03$\pm$3.87  &  $-$0.52$\pm$1.86  \\
20190125 & NT041453N482433N01 &  $-$0.62$\pm$2.77 &     0.34$\pm$1.24  &     0.46$\pm$1.31  &     0.03$\pm$1.47  &  $-$0.32$\pm$1.41  &  $-$0.17$\pm$2.46  &  $-$0.42$\pm$1.35  \\
20190126 & NT054421N274353N01 &  $-$0.55$\pm$3.06 &     0.46$\pm$1.25  &     0.46$\pm$1.41  &  $-$0.15$\pm$1.48  &  $-$0.38$\pm$1.52  &  $-$0.21$\pm$2.97  &  $-$0.56$\pm$1.42  \\
20190126 & NT054421N274353N02 &  $-$0.61$\pm$3.27 &     0.44$\pm$1.31  &     0.37$\pm$1.67  &  $-$0.21$\pm$1.55  &  $-$0.43$\pm$1.65  &  $-$0.33$\pm$3.23  &  $-$0.68$\pm$1.48  \\
		\hline
		\end{tabular}
\end{table*}

\begin{figure}
    \centering
	\includegraphics[angle=0,width=130mm]{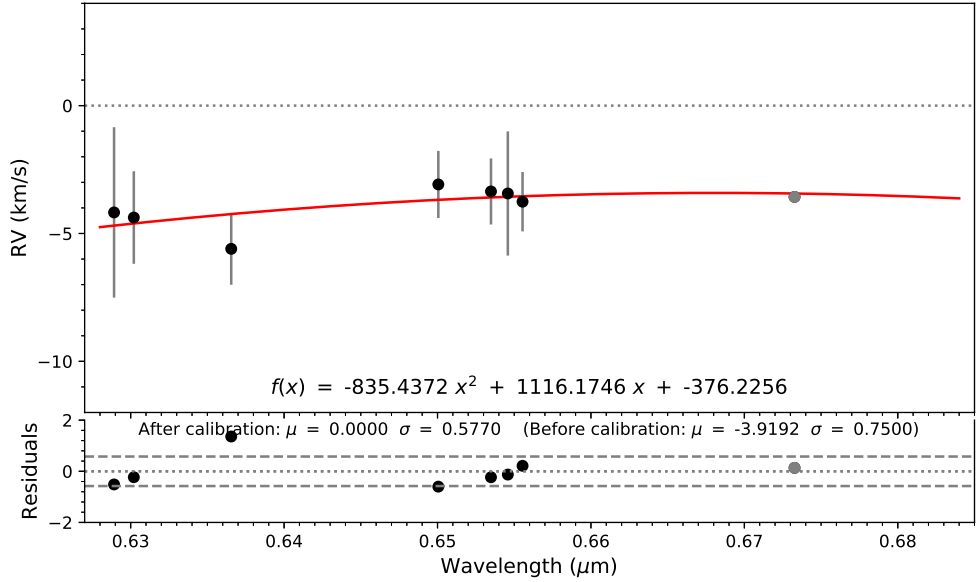}
    \caption{An example of the fitted RVs calibration function (plateID ``20180305-HD474690801"). The 7 black dots show the fitted mean RVs of the 7 single sky emission lines presented in Table\,\ref{tab:fit_results}, and the rightmost gray dot shows the extrapolated value at 6731\AA. The red curve plots the fitted RVs calibration function for this plate. The bottom panel shows the residuals, where the gray dotted horizontal line show the mean value of the residual and the gray dashed lines plot the 1$\sigma$ standard deviations. }
    \label{fig:fit_calibration}
\end{figure}

Then by using the fitted line center of these 7 single sky emission lines (as shown in Figure\,\ref{fig:hist_sky_RVs} and Table\,\ref{tab:rv_hist_results}), a second-order polynomial was used to do the fit. We also try the third- and fourth-order polynomial fit, but they show obvious over-fitting, thus we suggest that the second-order polynomial fitting is the best choice. Furthermore, as we mentioned before, the 7 single sky emission lines we used cover the wavelength range: 6290\,--\,6560\,\AA, which should be enough to calibrate the RVs of \Ha\ and \Lines{N}{II}{6548,\,6584} emission lines of nebulae. While another important emission line, i.e. \Lines{S}{II}{6717,\,6731} locates far from this range. We then extrapolate the RVs calibration to 6731\,\AA\ by applying a constraint that the RVs calibration at 6731\,\AA\ and 6548\,\AA\ are similar. Thus at least we can correct the systematic deviations at $\sim$\,6731\,\AA. Finally the RVs calibration function is fitted by using the 7 single sky emission lines and the value extrapolated at 6731\,\AA. Figure\,\ref{fig:fit_calibration} shows an example of the fitted RVs calibration function (the red curve). 

After the calibration, the sky RVs can go to as accurate as 0\,\kms (which is the true RVs of sky emission lines) with a standard deviation 0.5770\,\kms, while before the calibration, the sky RVs have a systemic uncertainty as large as $-$3.9192\,\kms\ with a standard deviation 0.7500\,\kms, for plate``20180305-HD474690801" as shown in Figure\,\ref{fig:fit_calibration}. Furthermore, when fitting the RVs calibration function, we suggest to do a RV error cut 3\,\kms\ of the sky emission lines. We also try a RV error cut 1.5\,\kms, the fitted RVs calibration function did not show large difference, thus the RV error cut of 3\,\kms\ is suggested to be used.

\begin{figure}
    \centering
	\includegraphics[angle=0,width=155mm]{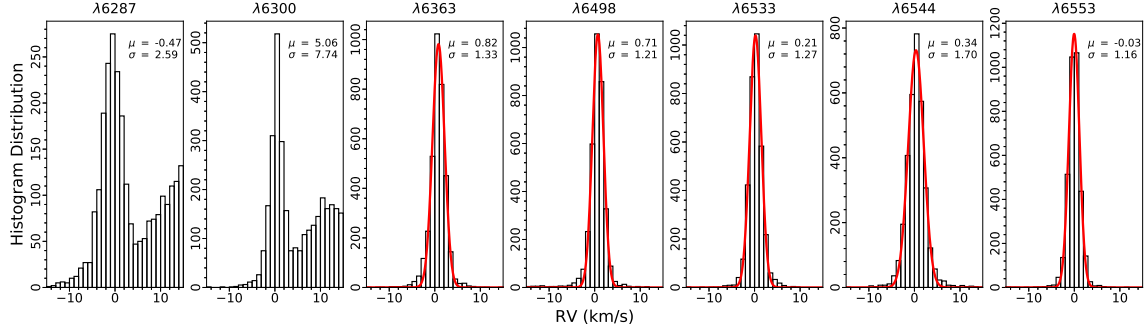}
    \caption{Same as Figure\,\ref{fig:hist_sky_RVs}, but for plate ``20181118-NT070341S003325N01", of which the distribution of sky RVs at $\lambda$6287 and $\lambda$6300 have multi-peaks, thus are not used for fitting the RVs calibration function of this plate.}
    \label{fig:hist_sky_RVs_20181118-NT070341S003325N01}
\end{figure}

\begin{table*}
	\centering
	\caption{The parameters of the RVs calibration function (f($x$)\,=\,a\,$x^2$\,+\,b$x$\,+c, where $x$ is the wavelength in unit of $\mu$m) for 15 MRS-N plates. The mean and standard deviation of sky RVs before and after the calibration are also listed. The last column ``Flag" marks the sky lines (i.e. the line number as shown in the first column of Table\,\ref{tab:sky_lines_single}) used to fit the RVs calibration function. }
	\scriptsize
	\label{tab:fit_results}
	\setlength{\tabcolsep}{3pt}
	\begin{tabular}{ccrrrrrrrc} 
		\hline
		 ObsDate & plateID & a & b & c & $\mu_\mathrm{before}$ & $\sigma_\mathrm{before}$ & $\mu_\mathrm{after}$ &  $\sigma_\mathrm{after}$ & Flag \\
		     &    &   &   &   & (\kms)& (\kms) & (\kms)& (\kms) & \\
		\hline
20180305 & HD474690101        &  $-$738.7490 &     989.4288 &  $-$334.8446 & $-$4.0503 & 0.7568 & 0.0000 & 0.6041 & 1, 2, 3, 4, 5, 6, 7 \\
20180305 & HD474690801        &  $-$835.4372 &    1116.1746 &  $-$376.2256 & $-$3.9192 & 0.7500 & 0.0000 & 0.5770 & 1, 2, 3, 4, 5, 6, 7 \\
20181018 & NT054437N290059N01 & $-$5739.5264 &    7601.2190 & $-$2518.0593 & $-$3.1091 & 2.0729 & 0.0000 & 0.4036 & 2, 3, 4, 5, 6, 7 \\
20181118 & NT070341S003325N01 &     742.0199 &  $-$991.7874 &     331.5405 &    0.3730 & 0.3019 & 0.0000 & 0.1868 & 3, 4, 5, 6, 7 \\
20181128 & NT230608N631246N01 &  $-$135.3334 &     182.7665 &   $-$61.9151 & $-$0.2862 & 0.1686 & 0.0000 & 0.1570 & 3, 4, 5, 6, 7 \\
20181128 & NT235302N592517N01 &     118.2756 &  $-$154.3496 &      50.2610 & $-$0.0816 & 0.1334 & 0.0000 & 0.1319 & 3, 4, 5, 6, 7 \\
20181129 & NT010552N655815N01 &    1153.0985 & $-$1519.0363 &     500.4621 &    0.3454 & 0.2697 & 0.0000 & 0.1834 & 3, 4, 5, 6, 7 \\
20181129 & NT232020N601629N01 &     571.4184 &  $-$753.3775 &     248.4534 &    0.2159 & 0.1444 & 0.0000 & 0.1032 & 3, 4, 5, 6, 7  \\
20181216 & NT065450N031415N01 &  $-$425.1480 &     547.7008 &  $-$176.2284 &    0.0767 & 0.3814 & 0.0000 & 0.3664 & 1, 2, 3, 4, 5, 6, 7 \\
20181228 & NT023239N614403N01 &  $-$663.7146 &     864.6065 &  $-$281.2221 &    0.2142 & 0.3820 & 0.0000 & 0.3537 & 1, 2, 3, 4, 5, 6, 7 \\
20190124 & NT024709N603414N01 &  $-$311.3390 &     408.5562 &  $-$134.5974 & $-$0.6480 & 0.3334 & 0.0000 & 0.3211 & 1, 2, 3, 4, 5, 6, 7 \\
20190125 & NT041422N543127N01 &      69.5751 &   $-$99.9254 &      35.4064 & $-$0.1149 & 0.3594 & 0.0000 & 0.3327 & 1, 2, 3, 4, 5, 6, 7 \\
20190125 & NT041453N482433N01 &  $-$158.2488 &     197.4298 &   $-$61.5766 & $-$0.1214 & 0.3502 & 0.0000 & 0.3300 & 1, 2, 3, 4, 5, 6, 7 \\
20190126 & NT054421N274353N01 &      58.2406 &   $-$88.3461 &      32.6192 & $-$0.1598 & 0.3824 & 0.0000 & 0.3375 & 1, 2, 3, 4, 5, 6, 7 \\
20190126 & NT054421N274353N02 &      68.5436 &  $-$103.3951 &      37.9632 & $-$0.2390 & 0.3970 & 0.0000 & 0.3409 & 1, 2, 3, 4, 5, 6, 7 \\
		\hline
		\end{tabular}
\end{table*}

\section{Discussion}

We applied the method described in the previous section for all the 15 MRS-N plates. Table\,\ref{tab:fit_results} lists the fitted parameters of the RVs calibration function (f($x$)\,=\,a\,$x^2$\,+\,b$x$\,+c, where $x$ is the wavelength in unit of $\mu$m). We need to note that for the 10 plates observed in 2018 March/December and 2019 January, the 7 single sky emission lines (the black points shown in Figure\,\ref{fig:fit_calibration}) are already very good to provide accurate RVs calibration. 

\begin{figure}
    \centering
	\includegraphics[angle=0,width=155mm]{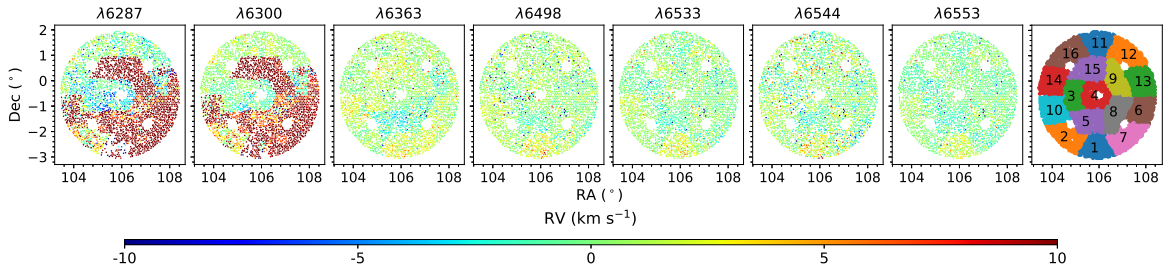}
    \caption{The coordinate (RA, Dec) distribution of the RVs of 7 sky emission lines for plate ``20181118-NT070341S003325N01". The color bar shows the value of sky RVs. For reference, the rightmost panel shows the distribution of the 16 spectrographs.}
    \label{fig:map-RVs_20181118-NT070341S003325N01}
\end{figure}

\begin{figure}
    \centering
	\includegraphics[angle=0,width=155mm]{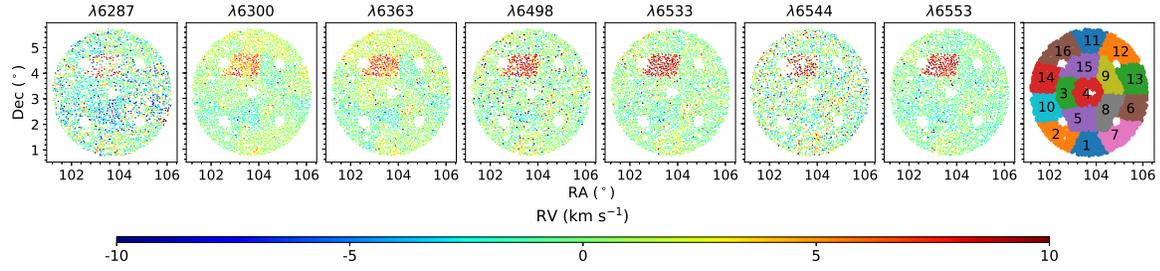}
    \caption{Same as Figure\,\ref{fig:map-RVs_20181118-NT070341S003325N01}, but for plate ``20181216-NT065450N031415N01".}
    \label{fig:map-RVs_20181216-NT065450N031415N01}
\end{figure}

\begin{figure}
    \centering
	\includegraphics[angle=0,width=155mm]{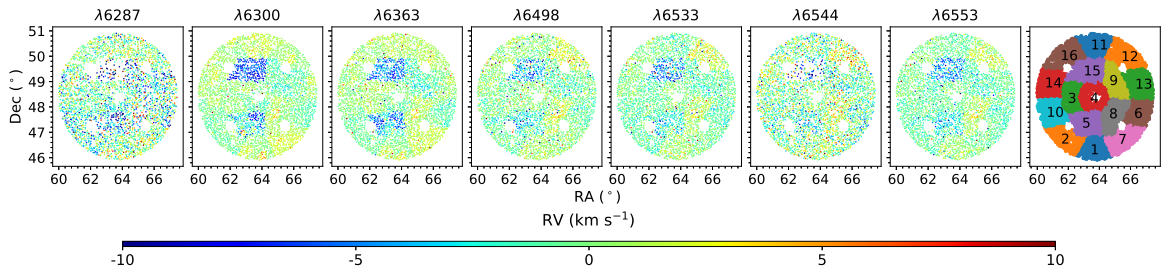}
    \caption{Same as Figure\,\ref{fig:map-RVs_20181118-NT070341S003325N01}, but for plate ``20190125-NT041453N482433N01".}
    \label{fig:map-RVs_20190125-NT041453N482433N01}
\end{figure}

\begin{figure}
    \centering
	\includegraphics[angle=0,width=155mm]{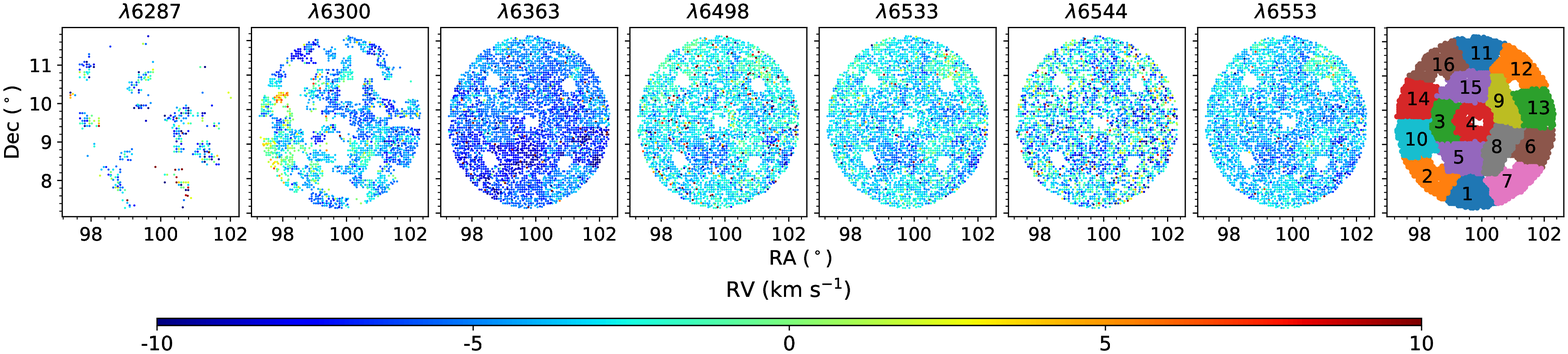}
    \caption{Same as Figure\,\ref{fig:map-RVs_20181118-NT070341S003325N01}, but for plate ``20180305-HD474690801".}
    \label{fig:map-RVs_20180305-HD474690801}
\end{figure}

But for the 5 plates observed in 2018 November, we need to remove two sky lines $\lambda$6287 and $\lambda$6300 in the blue end (for the one plate in 2018 October, only remove one sky line $\lambda$6287), as the histogram distribution of the line center of $\lambda$6287 and $\lambda$6300 are not single Gaussian symmetric or even multi-peaks (which is the situation shown in Figure\,\ref{fig:hist_sky_RVs_20181118-NT070341S003325N01}). In order to investigate the multi-peaks features of these sky RVs of the plates on 2018 November, we present the coordinate distribution of 7 sky RVs for plate ``20181118-NT070341S003325N01" in Figure\,\ref{fig:map-RVs_20181118-NT070341S003325N01}. For reference, the rightmost panel shows the spacial distribution of the 16 spectrographs. We can see that for the sky RVs from $\lambda$6287 and $\lambda$6300, almost half of the spectrographs focused around another peak (see the red dots in the left two panels). By carefully checking the spectra obtained in the ``green" and ``red" spectrographs (the blue and red dots in the left two panels), we can see exact and clear large shift of the emission line $\lambda$6287 and $\lambda$6300. That means, for these five MRS-N plates observed in 2018 November, there are large systematic deviations near 6300\AA\ for almost half of the spectrographs, which should be due to the bad wavelength calibration in this region. As we can see in the bottom panel of Figure\,\ref{fig:arc_lines}, there are only two very week arc lines near 6300\,\AA\, which thus may lead to bad wavelength calibration near this region when these two lines are too weak and have bad quality.

In December 2018, the systematic deviation near 6300\,\AA\ has disappeared. Figure\,\ref{fig:map-RVs_20181216-NT065450N031415N01} shows the coordinate distribution of sky RVs for one plate observed in December 2018, i.e. ``20181216-NT065450N031415N01". Although the multi-peaks problem has disappeared near 6300\,\AA, there is another obvious problem: one of the spectrograph, i.e spectrograph 15 has large ($\gtrsim$\,5\,\kms) systematic difference from other spectrographs. Then by checking the 5 plates observed in 2019 January, we find this effect still exists. Figure\,\ref{fig:map-RVs_20190125-NT041453N482433N01} shows the situation of one plate obtained in 2019 January, i.e. plate ``20190125-NT041453N482433N01", from which we can see that the systematic difference of spectrograph becomes minus by comparing with Figure\,\ref{fig:map-RVs_20181216-NT065450N031415N01}, and another spectrograph (ID: 5) has the same situation. This can be explained due to the spectrograph drift for these two spectrographs (i.e. 15 and 5).

While for the two plates observed in March 2018, the situation is very different from the above plates (see Figure\,\ref{fig:map-RVs_20180305-HD474690801}), which is totally dominated by the systematic deviation ($\sim$\,4\,\kms, which can also be seen from Table\,\ref{tab:fit_results}) of the whole spectra for the whole plate. And those bluer than 6363\,\AA\ are even worse that those redder than 6363\,\AA, which can also be seen from Figure\,\ref{fig:fit_calibration} and \ref{fig:hist_sky_RVs}. This can be easily explained, as in March 2018, the arc lamp MRS used is Sc lamp, which only has very limited number of arc lines (as shown in the top panel of Figure\,\ref{fig:arc_lines}), thus affected the wavelength calibration and finally resulted in the large systematic deviations of the two MRS-N plates in March 2018. 

From Table\,\ref{tab:rv_hist_results}\,--\,\ref{tab:fit_results} and above discussions, we can see that for the two MRS-N plates in 2018 March and one plate in October, there are large systematic deviations around 2\,$\sim$\,4\,\kms\ for the whole plate, which can be roughly eliminated after our calibration presented in Table\,\ref{tab:fit_results}. While for the five plates observed in 2018 November, two plates in 2018 December, and five plates in 2019 January, the systematic deviations of the whole plates are around 0.2\,$\sim$\,0.5\,\kms, which can also be roughly calibrated after applying our calibrations in Table\,\ref{tab:fit_results}. The decreasing of the systematic deviations of the whole plates from early 2018 to late 2018 indicates that the wavelength calibration has been improved at least since 2018 November. The improvement of MRS wavelength calibration from early 2018 to 2019 was mainly due to the updates of arc lamps: (1) At early 2018, the arc lamp used by MRS was Sc which only had a very limited number of available arc lines as we mentioned before, and the Sc lamp had serious systematic deviation about 4\,$\sim$\,5\,\kms; (2) From May 2018, MRS began to use Th-Ar lamp, but only 10 Th-Ar lamps were used initially. Due to the long optical path and large focal plane of LAMOST, only using 10 Th-Ar lamps were not enough to get high enough brightness and uniform distribution of illumination, thus may affect the wavelength calibration. (3) At the end of 2018 and beginning of 2019, the number of Th-Ar lamps has been increased from 10 to 20, which enhanced the brightness and uniformity of illumination, thus improved the wavelength calibration of MRS in 2019.

\begin{table*}
	\centering
	\caption{The parameters of the RVs calibration function (f($x$)\,=\,a\,$x^2$\,+\,b$x$\,+c, where $x$ is the wavelength in unit of $\mu$m) for the 16 spectrographs of 15 MRS-N plates. The mean and standard deviation of sky RVs before and after the calibration are also listed. The last column ``Flag" marks the sky lines (i.e. the line number as shown in the first column of Table\,\ref{tab:sky_lines_single}) used to fit the RVs calibration function. The complete table can be found online. }
	\scriptsize
	\label{tab:fit_results_each_sp}
	\setlength{\tabcolsep}{3pt}
	\begin{tabular}{ccrrrrrrrc} 
		\hline
		 Spectrograph ID & a & b & c & $\mu_\mathrm{before}$ & $\sigma_\mathrm{before}$ & $\mu_\mathrm{after}$ &  $\sigma_\mathrm{after}$ & Flag \\
		     &    &   &   &   & (\kms)& (\kms) & (\kms)& (\kms) & \\
		\hline
20180305-HD474690101-01 & $-$2456.7311 &    3270.5386 & $-$1091.5202 & $-$4.0119 & 1.1733 & 0.0000 & 0.5054 & 2, 3, 4, 5, 6, 7 \\
20180305-HD474690101-02 &   $-$32.0486 &      77.3790 &   $-$40.8026 & $-$4.0333 & 1.3238 & 0.0000 & 1.2405 & 2, 3, 4, 5, 6, 7 \\
20180305-HD474690101-03 & $-$1273.8805 &    1693.2619 &  $-$566.2647 & $-$4.0538 & 0.6751 & 0.0000 & 0.4312 & 2, 3, 4, 5, 6, 7 \\
20180305-HD474690101-04 & $-$2166.7051 &    2870.7681 &  $-$954.6600 & $-$4.6506 & 1.0882 & 0.0000 & 0.5559 & 1, 2, 3, 4, 5, 6, 7 \\
20180305-HD474690101-05 &    1220.1428 & $-$1580.3003 &     506.3375 & $-$5.1127 & 1.1884 & 0.0000 & 1.1602 & 1, 2, 3, 4, 5, 6, 7 \\
20180305-HD474690101-06 &  $-$566.5774 &     764.2089 &  $-$261.8449 & $-$4.6631 & 0.9684 & 0.0000 & 0.8718 & 1, 2, 3, 4, 5, 6, 7 \\
20180305-HD474690101-07 & $-$1200.3801 &    1568.7877 &  $-$516.6348 & $-$4.2810 & 0.5673 & 0.0000 & 0.5080 & 2, 3, 4, 5, 6, 7 \\
20180305-HD474690101-08 &     240.1561 &  $-$287.3387 &      81.5520 & $-$3.7576 & 1.0876 & 0.0000 & 1.0308 & 1, 2, 3, 4, 5, 6, 7 \\
20180305-HD474690101-09 &  $-$587.3293 &     795.5451 &  $-$273.2511 & $-$4.4821 & 0.8819 & 0.0000 & 0.7382 & 1, 2, 3, 4, 5, 6, 7 \\
20180305-HD474690101-10 &     261.0949 &  $-$327.5368 &      98.7891 & $-$3.7470 & 1.1004 & 0.0000 & 1.0875 & 2, 3, 4, 5, 6, 7 \\
20180305-HD474690101-11 & $-$1329.2261 &    1757.0361 &  $-$584.1550 & $-$3.8879 & 0.5716 & 0.0000 & 0.3699 & 2, 3, 4, 5, 6, 7 \\
20180305-HD474690101-12 & $-$3540.2103 &    4693.6972 & $-$1557.8253 & $-$3.2085 & 1.3672 & 0.0000 & 0.3912 & 2, 3, 4, 5, 6, 7 \\
20180305-HD474690101-13 & $-$2064.6018 &    2722.2270 &  $-$900.7025 & $-$3.8765 & 0.6929 & 0.0000 & 0.3311 & 2, 3, 4, 5, 6, 7 \\
20180305-HD474690101-14 &     797.1745 & $-$1044.9741 &     339.1450 & $-$3.1014 & 0.9919 & 0.0000 & 0.9674 & 1, 2, 3, 4, 5, 6, 7 \\
20180305-HD474690101-15 &  $-$309.2930 &     426.8453 &  $-$150.7229 & $-$4.0662 & 0.7151 & 0.0000 & 0.6162 & 1, 2, 3, 4, 5, 6, 7 \\
20180305-HD474690101-16 & $-$4376.8213 &    5796.9788 & $-$1921.7888 & $-$3.6397 & 1.6030 & 0.0000 & 0.3867 & 2, 3, 4, 5, 6, 7 \\
20180305-HD474690801-01 & $-$2484.6439 &    3296.8155 & $-$1096.5995 & $-$3.8144 & 1.1004 & 0.0000 & 0.5585 & 2, 3, 4, 5, 6, 7 \\
20180305-HD474690801-02 &     214.8080 &  $-$252.7951 &      69.5716 & $-$4.0027 & 1.2683 & 0.0000 & 1.2134 & 1, 2, 3, 4, 5, 6, 7 \\
20180305-HD474690801-03 & $-$1340.9390 &    1775.6265 &  $-$591.4048 & $-$4.0029 & 0.6132 & 0.0000 & 0.3912 & 2, 3, 4, 5, 6, 7 \\
20180305-HD474690801-04 & $-$1995.4816 &    2654.7405 &  $-$886.3688 & $-$4.4149 & 1.1345 & 0.0000 & 0.5393 & 1, 2, 3, 4, 5, 6, 7 \\
		\hline
		\end{tabular}
\end{table*}

\begin{figure}
    \centering
	\includegraphics[angle=0,width=155mm]{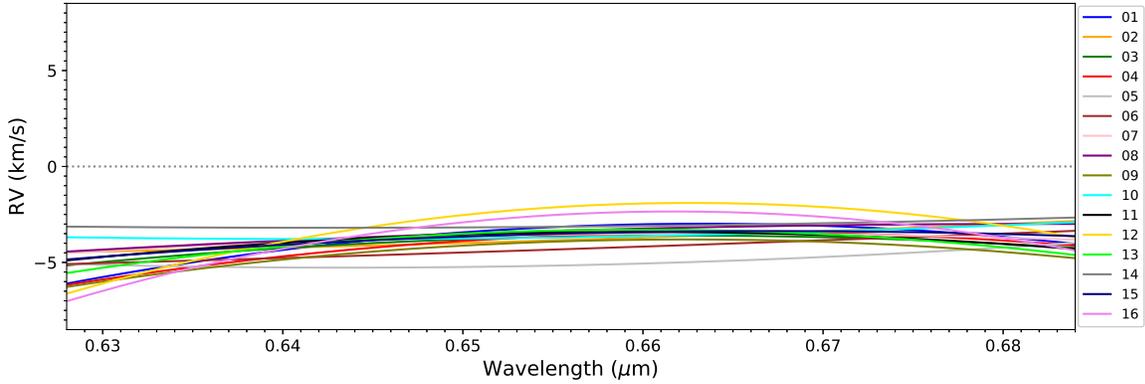}
    \caption{The fitted RVs calibration function of 16 spectrographs of plate ``20180305-HD474690801". The colors show different spectrographs.}
    \label{fig:colored-sp_20180305-HD474690801}
\end{figure}

\begin{figure}
    \centering
	\includegraphics[angle=0,width=155mm]{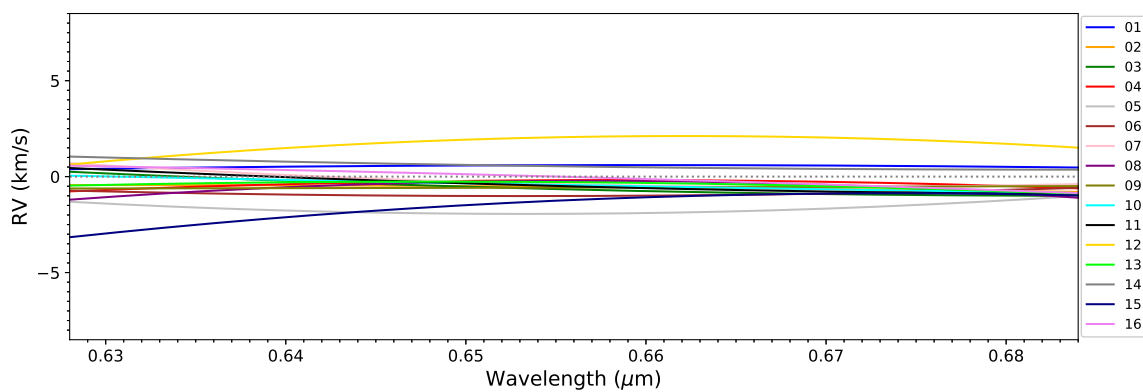}
    \caption{Same as Figure\,\ref{fig:colored-sp_20180305-HD474690801}, but for the plate ``20190126-NT054421N274353N01".}
    \label{fig:colored-sp_20190126-NT054421N274353N01}
\end{figure}

Although the wavelength calibration of MRS has been improved at the end of 2018 and beginning of 2019, there still some problems such as the spectrograph drift effect as mentioned previously (see Figure\,\ref{fig:map-RVs_20181216-NT065450N031415N01}\,--\,\ref{fig:map-RVs_20190125-NT041453N482433N01}. Thus, we also investigate the distribution of sky RVs for different spectrographs of the same plate. Figure\,\ref{fig:rv-hist_sp-04} and \ref{fig:rv-hist_sp-08} in the appendix shows the histogram distribution of sky RVs and the fitted calibration function for two different spectrograph (i.e. spectrograph ID 04 and 08 of plate ``20180305-HD474690801") of one plate. We can see clear difference between the systematic deviation (around 0.5\,\kms) of the these two spectrographs. Figure\,\ref{fig:colored-sp_20180305-HD474690801} plots the RVs calibration function of all the 16 spectrographs of plate ``20180305-HD474690801". There is a systematic deviation about 3\,$\sim$\,4\,\kms\ between these 16 spectrographs, which implies the RVs calibration should be different for different spectrographs even in a same plate. Even for those plates in January 2019, the systematic deviations between different spectrographs still exist, as shown in Figure\,\ref{fig:colored-sp_20190126-NT054421N274353N01} and Figure\,\ref{fig:map-RVs_20190125-NT041453N482433N01}. Table\,\ref{tab:fit_results_each_sp} list the fitted calibration function for the 16 spectrographs of the 15 MRS-N plates.

\begin{figure}
    \centering
	\includegraphics[angle=0,width=155mm]{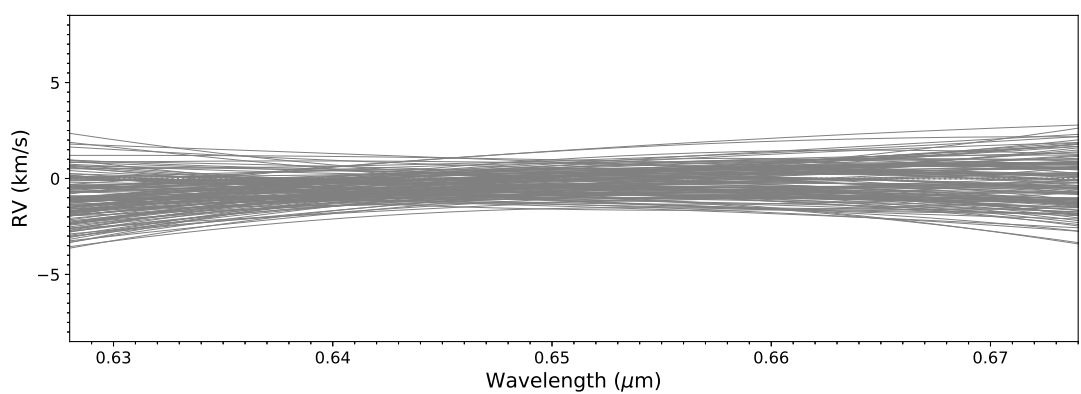}
    \caption{The fitted RVs calibration function of all the spectra in one spectrograph (ID: 04) of plate ``20190126-NT054421N274353N02".}
    \label{fig:calibration_260spec}
\end{figure}

\begin{figure}
    \centering
	\includegraphics[angle=0,width=155mm]{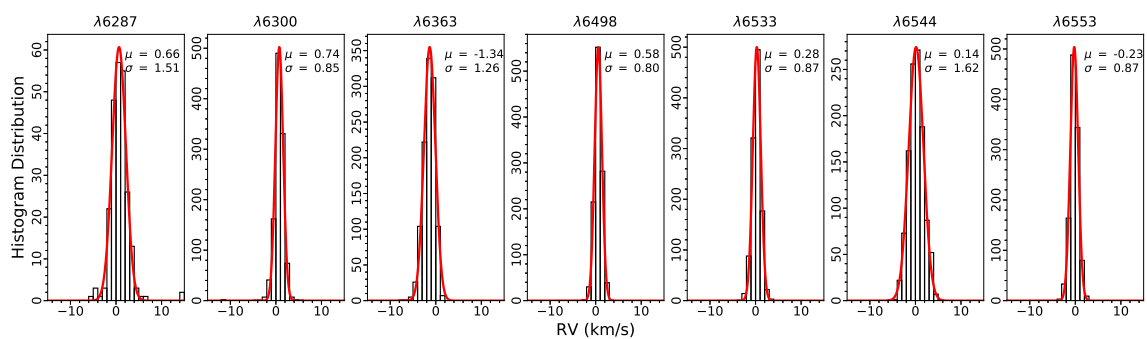}
    \caption{The histogram distribution of sky RVs after applying the RVs calibration function in Table\,\ref{tab:fit_results_each_spec} (for plate ``20180305-HD474690801").}
    \label{fig:hist_sky_RVs_after-calibration}
\end{figure}

\begin{table*}
	\centering
	\caption{The parameters of the RVs calibration function (f($x$)\,=\,a\,$x^2$\,+\,b$x$\,+c, where $x$ is the wavelength in unit of $\mu$m) for every spectrum of the 15 MRS-N plates. The last column ``Flag" marks the sky lines (i.e. the line number as shown in the first column of Table\,\ref{tab:sky_lines_single}) we used to fit the RVs calibration function. The complete table can be found online.}
	\scriptsize
	\label{tab:fit_results_each_spec}
	\setlength{\tabcolsep}{1.5pt}
	\begin{tabular}{crrrrrrrc} 
		\hline
		SpecID & a & b & c & $\mu_\mathrm{before}$ & $\sigma_\mathrm{before}$ & $\mu_\mathrm{after}$ &  $\sigma_\mathrm{after}$ & Flag \\
		       &  &   &  & (\kms) & (\kms) & (\kms) & (\kms) &  \\
		\hline
20180305-HD474690101-01-007 & $-$2648.5858 &  3452.2209 & $-$1127.7720 & $-$3.2969 & 1.2592 & 0.0000 & 1.1414 & 2, 3, 4, 5, 6, 7 \\
20180305-HD474690101-01-008 & $-$1607.8895 &  2142.9616 &  $-$715.9577 & $-$2.6088 & 0.9337 & 0.0000 & 0.5929 & 2, 3, 4, 5, 6, 7 \\
20180305-HD474690101-01-013 & $-$1700.5175 &  2243.1382 &  $-$742.9757 & $-$3.6736 & 0.9163 & 0.0000 & 0.7608 & 2, 3, 4, 5, 6, 7 \\
20180305-HD474690101-01-016 & $-$2461.9163 &  3284.7194 & $-$1097.8069 & $-$3.2688 & 1.3906 & 0.0000 & 0.7883 & 2, 3, 4, 5, 6, 7 \\
20180305-HD474690101-01-017 & $-$1924.3222 &  2584.4421 &  $-$869.8695 & $-$3.2884 & 1.4040 & 0.0000 & 0.8745 & 2, 3, 4, 5, 6, 7 \\
20180305-HD474690101-01-018 & $-$3428.3998 &  4525.5355 & $-$1496.2565 & $-$3.6981 & 1.2518 & 0.0000 & 0.6652 & 2, 3, 4, 5, 6, 7 \\
20180305-HD474690101-01-019 & $-$1635.4571 &  2246.1514 &  $-$771.9583 & $-$3.1524 & 1.8688 & 0.0000 & 1.0396 & 2, 3, 4, 5, 6, 7 \\
20180305-HD474690101-01-020 & $-$3727.5943 &  4977.1512 & $-$1662.7970 & $-$3.1154 & 1.9093 & 0.0000 & 0.6945 & 2, 3, 4, 5, 6, 7 \\
20180305-HD474690101-01-022 & $-$4079.3504 &  5453.0305 & $-$1823.4475 & $-$3.1107 & 2.1314 & 0.0000 & 0.6809 & 2, 3, 4, 5, 6, 7 \\
20180305-HD474690101-01-024 & $-$4138.4885 &  5465.6417 & $-$1808.0232 & $-$4.5188 & 1.4273 & 0.0000 & 0.5717 & 2, 3, 4, 5, 6, 7 \\
20180305-HD474690101-01-025 & $-$4198.3574 &  5553.6022 & $-$1839.4810 & $-$4.1017 & 1.4810 & 0.0000 & 0.4275 & 2, 3, 4, 5, 6, 7 \\
20180305-HD474690101-01-027 & $-$1727.4051 &  2253.5927 &  $-$739.1745 & $-$4.4544 & 0.9819 & 0.0000 & 0.9172 & 2, 3, 4, 5, 6, 7 \\
		\hline
		\end{tabular}
\end{table*}

Although most of the sky RVs from one plate or spectrograph are focused near a peak (as shown in Figure\,\ref{fig:hist_sky_RVs}, \ref{fig:rv-hist_sp-04} and \ref{fig:rv-hist_sp-08}), there are still some fibers with sky RVs a little far from the peak value. To obtain higher precision of RV determinations, it's also necessary to provide a RV calibration function for every nebula spectrum, not only the whole plate or spectrograph. Figure\,\ref{fig:calibration_260spec} shows an example of the fitted RVs calibration functions of all the spectra in one spectrograph of one plate. We can see that even in one spectrograph, the fitted calibration function of each spectrum is still a little different. Thus it should be better to do the calibration for each spectrum. Table\,\ref{tab:fit_results_each_spec} lists the parameters of the fitted RVs calibration function for every spectrum. For those without sky RVs available due to their low spectral quality or shortage of data, the RV calibration function of very nearby spectra can be applied instead. Figure\,\ref{fig:hist_sky_RVs_after-calibration} shows an example (for plate ``20180305-HD474690801") of the histogram distribution of sky RVs after applying the RVs calibration function provided in Table\,\ref{tab:fit_results_each_spec}. We can see that the systematic deviation as large as 4\,\kms\ has disappeared, especially for those sky lines near 6550\AA\ which harboring the important nebulae emission features (i.e. \Ha, \Lines{N}{II}{6548,\,6584}).

The future MRS-N pipeline (Wu et al., 2020c, in preparation) will adopt the method developed here to investigate the precision of wavelength calibration of MRS-N nebulae spectra in real time, and fit a RVs calibration function for every nebula spectrum by using sky emission lines to finally improve the precision of nebulae RVs. Furthermore, this method is also valuable for other sub-surveys of MRS if they have some plates fortunately with available strong sky emission lines.

\section{Summary}
LAMOST\,II MRS survey has initiated a sub-survey of nebulae, i.e. MRS-N survey since 2017 September. MRS-N survey will monitor a large sample of nebulae emission line features, mainly including \Lines{N}{II}{6548,\,6584}, \Ha, and \Lines{S}{II}{6717,\,6731}. Before studying the kinematics and dynamics properties of nebulae, it's important to investigating the precision of the wavelength calibration of nebulae spectra, and finally calibrate RVs of nebulae emission lines. 

We propose a method to investigate the precision of wavelength calibration of MRS-N, by using the sky emission lines in the red spectra of MRS. We find that the RVs of the 15 observed MRS-N plates have systematic deviations from $\sim$\,0.2 to 4\,\kms for different plates obtained at different times. A RVs calibration function is fitted by using the fitted centers of the sky emission lines, which can successfully calibrate the RVs deviations of nebulae in real time. The future released value-added catalogue of MRS-N will apply this RVs calibration function to provide accurate RVs of nebulae emission lines. We suggest other MRS-N users to use our method to investigate and improve the RVs precision of nebulae spectra in the future too.
\normalem

\begin{acknowledgements}
    This project is supported by the National Key R \& D Program of China (No. 2017YFA0402704) and National Natural Science Foundation of China (grant Nos. 11903048, 11833006, U1531244, U1831209, 11733006, 11403061, U1531118, 11973060, U1631131, 11873057).  C.-H. Hsia acknowledges the supports from The Science and Technology Development Fund, Macau SAR (file no. 061/2017/A2 and 0007/2019/A) and Faculty Research Grants of the Macau University of Science and Technology (program no. FRG-19-004-SSI). 
    
    Guoshoujing Telescope (the Large Sky Area Multi-Object Fiber Spectroscopic Telescope LAMOST) is a National Major Scientific Project built by the Chinese Academy of Sciences. Funding for the project has been provided by the National Development and Reform Commission. LAMOST is operated and managed by the National Astronomical Observatories, Chinese Academy of Sciences. Supported by Key Research Program of Frontier Sciences, CAS, Grant NO. QYZDY-SSW-SLH007.

\end{acknowledgements}

\bibliographystyle{raa}
\bibliography{ms2020-0128}

\begin{thebibliography}{34}
\providecommand\natexlab[1]{#1}
\providecommand\JournalTitle[1]{#1}

\bibitem[{Alvarez} \& {Hoare}(2005)]{Alvarez2005A&A...440..569A}
{Alvarez}, C., \& {Hoare}, M.~G. 2005, \aap, 440, 569

\bibitem[{Blair} {et~al.}(1991)]{Blair1991ApJ...366..484B}
{Blair}, W.~P., {Long}, K.~S., \& {Vancura}, O. 1991, \apj, 366, 484

\bibitem[{Cavichia} {et~al.}(2017)]{Cavichia2017MNRAS.468..272C}
{Cavichia}, O., {Costa}, R.~D.~D., {Maciel}, W.~J., \& {Moll{\'a}}, M. 2017,
  \mnras, 468, 272

\bibitem[{Chen} {et~al.}(2017)]{Chen2017MNRAS.472.3924C}
{Chen}, B.~Q., {Liu}, X.~W., {Ren}, J.~J., {et~al.} 2017, \mnras, 472, 3924

\bibitem[{Cui} {et~al.}(2012)]{Cui2012RAA....12.1197C}
{Cui}, X.-Q., {Zhao}, Y.-H., {Chu}, Y.-Q., {et~al.} 2012, Research in Astronomy
  and Astrophysics, 12, 1197

\bibitem[{Damiani} {et~al.}(2016)]{Damiani2016A&A...591A..74D}
{Damiani}, F., {Bonito}, R., {Magrini}, L., {et~al.} 2016, \aap, 591, A74

\bibitem[{Deng} {et~al.}(2012)]{Deng2012RAA....12..735D}
{Deng}, L.-C., {Newberg}, H.~J., {Liu}, C., {et~al.} 2012, Research in
  Astronomy and Astrophysics, 12, 735

\bibitem[{Esteban} {et~al.}(2019)]{Esteban2019arXiv190510129E}
{Esteban}, C., {Garc{\'\i}a-Rojas}, J., {Arellano-C{\'o}rdova}, K.~Z., \&
  {M{\'e}ndez-Delgado}, J.~E. 2019, arXiv e-prints, arXiv:1905.10129

\bibitem[{Fesen} {et~al.}(1985)]{Fesen1985ApJ...292...29F}
{Fesen}, R.~A., {Blair}, W.~P., \& {Kirshner}, R.~P. 1985, \apj, 292, 29

\bibitem[{Gerardy} \& {Fesen}(2007)]{Gerardy2007MNRAS.376..929G}
{Gerardy}, C.~L., \& {Fesen}, R.~A. 2007, \mnras, 376, 929

\bibitem[{Kopsacheili} {et~al.}(2020)]{Kopsacheili2020MNRAS.491..889K}
{Kopsacheili}, M., {Zezas}, A., \& {Leonidaki}, I. 2020, \mnras, 491, 889

\bibitem[{Kwitter} {et~al.}(2012)]{Kwitter2012ApJ...753...12K}
{Kwitter}, K.~B., {Lehman}, E. M.~M., {Balick}, B., \& {Henry}, R.~B.~C. 2012,
  \apj, 753, 12

\bibitem[{Liu} {et~al.}(2020)]{Liu2020arXiv200507210L}
{Liu}, C., {Fu}, J., {Shi}, J., {et~al.} 2020, arXiv e-prints, arXiv:2005.07210

\bibitem[{Liu} {et~al.}(2019)]{Liu2019RAA....19...75L}
{Liu}, N., {Fu}, J.-N., {Zong}, W., {et~al.} 2019, Research in Astronomy and
  Astrophysics, 19, 075

\bibitem[{Luo} {et~al.}(2012)]{Luo2012RAA....12.1243L}
{Luo}, A.~L., {Zhang}, H.-T., {Zhao}, Y.-H., {et~al.} 2012, Research in
  Astronomy and Astrophysics, 12, 1243

\bibitem[{Luo} {et~al.}(2015)]{Luo2015RAA....15.1095L}
{Luo}, A.~L., {Zhao}, Y.-H., {Zhao}, G., {et~al.} 2015, Research in Astronomy
  and Astrophysics, 15, 1095

\bibitem[{Merrett} {et~al.}(2006)]{Merrett2006MNRAS.369..120M}
{Merrett}, H.~R., {Merrifield}, M.~R., {Douglas}, N.~G., {et~al.} 2006, \mnras,
  369, 120

\bibitem[{Osterbrock} \& {Ferland}(2006)]{Osterbrock2006agna.book.....O}
{Osterbrock}, D.~E., \& {Ferland}, G.~J. 2006, {Astrophysics of gaseous nebulae
  and active galactic nuclei}

\bibitem[{Osterbrock} {et~al.}(1996)]{Osterbrock1996PASP..108..277O}
{Osterbrock}, D.~E., {Fulbright}, J.~P., {Martel}, A.~R., {et~al.} 1996, \pasp,
  108, 277

\bibitem[{Rebassa-Mansergas} {et~al.}(2007)]{Rebassa2007MNRAS.382.1377R}
{Rebassa-Mansergas}, A., {G{\"a}nsicke}, B.~T., {Rodr{\'\i}guez-Gil}, P.,
  {Schreiber}, M.~R., \& {Koester}, D. 2007, \mnras, 382, 1377

\bibitem[{Rebassa-Mansergas} {et~al.}(2016)]{Rebassa2016MNRAS.458.3808R}
{Rebassa-Mansergas}, A., {Ren}, J.~J., {Parsons}, S.~G., {et~al.} 2016, \mnras,
  458, 3808

\bibitem[{Rebassa-Mansergas} {et~al.}(2017)]{Rebassa2017MNRAS.472.4193R}
{Rebassa-Mansergas}, A., {Ren}, J.~J., {Irawati}, P., {et~al.} 2017, \mnras,
  472, 4193

\bibitem[{Ren} {et~al.}(2018{\natexlab{a}})]{Ren2018MNRAS.477.4641R}
{Ren}, J.~J., {Rebassa-Mansergas}, A., {Parsons}, S.~G., {et~al.}
  2018{\natexlab{a}}, \mnras, 477, 4641

\bibitem[{Ren} {et~al.}(2014)]{Ren2014A&A...570A.107R}
{Ren}, J.~J., {Rebassa-Mansergas}, A., {Luo}, A.~L., {et~al.} 2014, \aap, 570,
  A107

\bibitem[{Ren} {et~al.}(2018{\natexlab{b}})]{Ren2018RAA....18..111R}
{Ren}, J.-J., {Liu}, X.-W., {Chen}, B.-Q., {et~al.} 2018{\natexlab{b}},
  Research in Astronomy and Astrophysics, 18, 111

\bibitem[{Ren} {et~al.}(2013)]{Ren2013AJ....146...82R}
{Ren}, J., {Luo}, A., {Li}, Y., {et~al.} 2013, \aj, 146, 82

\bibitem[{Shaver} {et~al.}(1983)]{Shaver1983MNRAS.204...53S}
{Shaver}, P.~A., {McGee}, R.~X., {Newton}, L.~M., {Danks}, A.~C., \&
  {Pottasch}, S.~R. 1983, \mnras, 204, 53

\bibitem[{Su} \& {Cui}(2004)]{Su2004ChJAA...4....1S}
{Su}, D.-Q., \& {Cui}, X.-Q. 2004, \cjaa, 4, 1

\bibitem[{Wang} {et~al.}(2019)]{Wang2019ApJS..244...27W}
{Wang}, R., {Luo}, A.~L., {Chen}, J.~J., {et~al.} 2019, \apjs, 244, 27

\bibitem[{Wang} {et~al.}(1996)]{Wang1996ApOpt..35.5155W}
{Wang}, S.-G., {Su}, D.-Q., {Chu}, Y.-Q., {Cui}, X., \& {Wang}, Y.-N. 1996,
  \ao, 35, 5155

\bibitem[{Woltjer}(1972)]{Woltjer1972ARA&A..10..129W}
{Woltjer}, L. 1972, \araa, 10, 129

\bibitem[{Wu} {et~al.}(2020{\natexlab{a}})]{Wu2020RAA....20...33W}
{Wu}, C.-J., {Wu}, H., {Hsia}, C.-H., {et~al.} 2020{\natexlab{a}}, Research in
  Astronomy and Astrophysics, 20, 033

\bibitem[{Wu} {et~al.}(2020{\natexlab{b}})]{Wu2020arXiv200705240W}
{Wu}, C.-J., {Wu}, H., {Zhang}, W., {et~al.} 2020{\natexlab{b}}, arXiv
  e-prints, arXiv:2007.05240

\bibitem[{Zhao} {et~al.}(2012)]{Zhao2012RAA....12..723Z}
{Zhao}, G., {Zhao}, Y.-H., {Chu}, Y.-Q., {Jing}, Y.-P., \& {Deng}, L.-C. 2012,
  Research in Astronomy and Astrophysics, 12, 723

\end{thebibliography}

\setcounter{table}{0}
\renewcommand{\thetable}{A\arabic{table}}
\setcounter{figure}{0}
\renewcommand{\thefigure}{A\arabic{figure}}

\begin{landscape}
\begin{figure}
    \centering
	\includegraphics[angle=0,width=230mm]{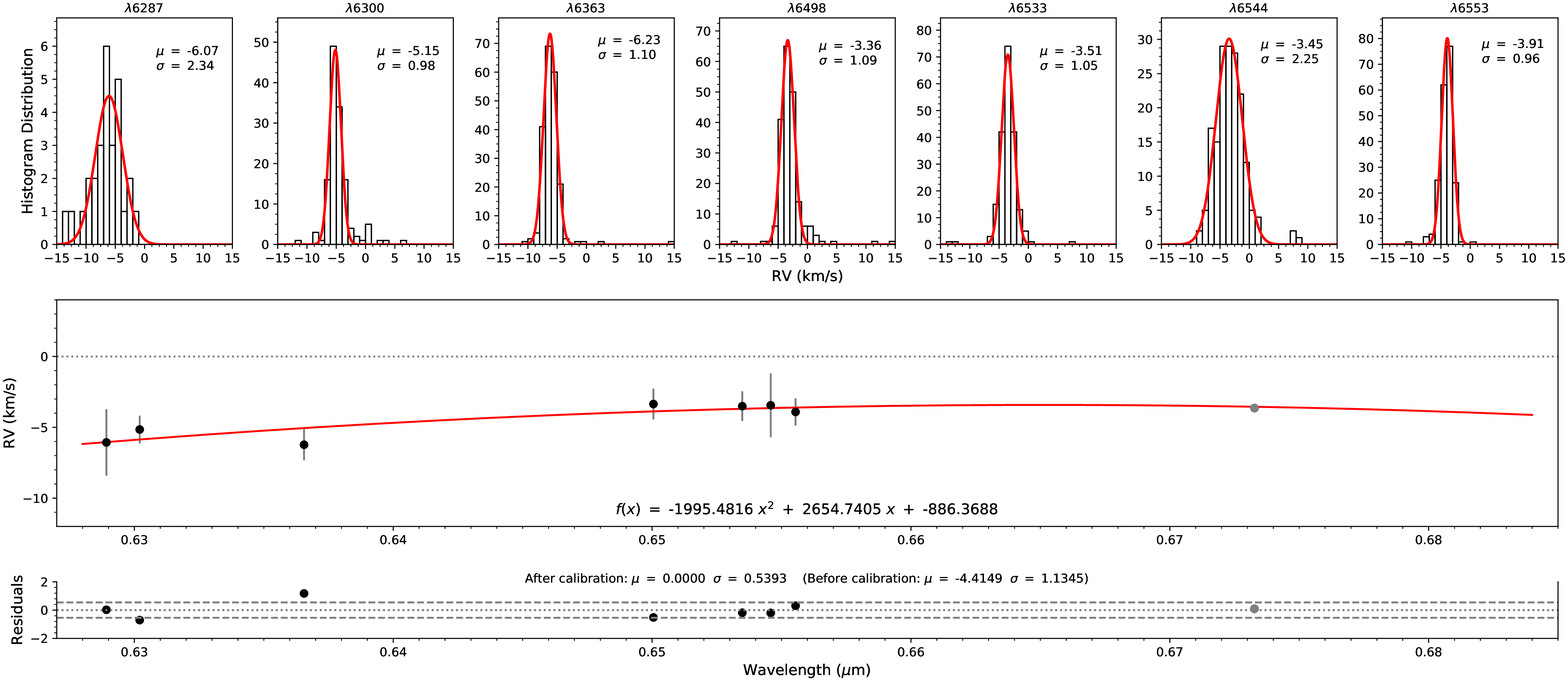}
    \caption{The upper panels: the histogram distribution of sky RVs for spectrograph ID 04 (i.e. ``20180305-HD474690801-04"); bottom panels: the corresponding fitted RVs calibration function for this spectrograph. }
    \label{fig:rv-hist_sp-04}
\end{figure}
\end{landscape}

\begin{landscape}
\begin{figure}
    \centering
	\includegraphics[angle=0,width=230mm]{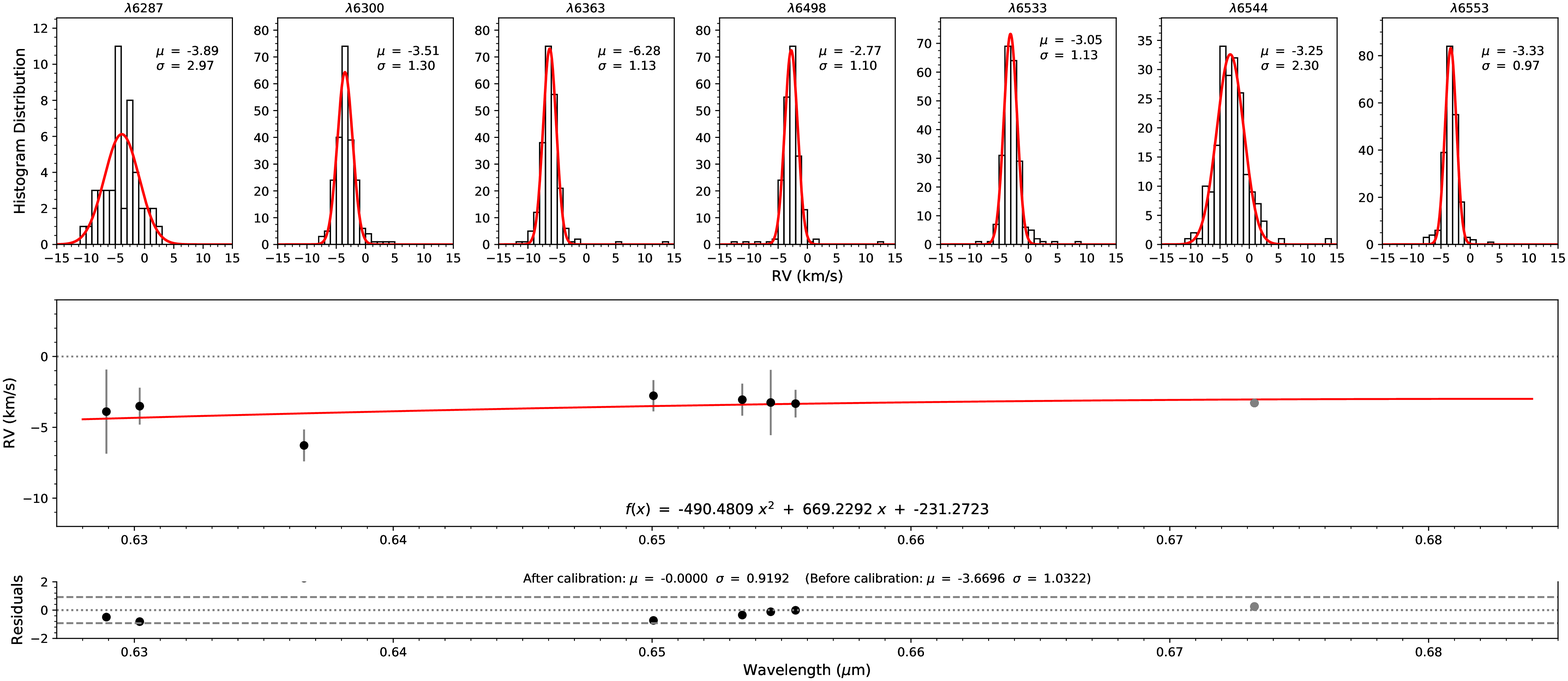}
    \caption{Same as Figure\,\ref{fig:rv-hist_sp-04}, but for spectrograph ID 08 (i.e. ``20180305-HD474690801-08").}
    \label{fig:rv-hist_sp-08}
\end{figure}
\end{landscape}

\end{document}